\documentclass[11pt,a4paper,preprintnumbers,nofootinbib]{revtex4}

\pdfoutput=1

\usepackage[T1]{fontenc}
\usepackage[utf8]{inputenc}
\usepackage{amsmath,amssymb,amsfonts,mathtools,bm}
\usepackage{hyperref}
\usepackage[final]{graphicx}
\usepackage{xcolor}
\usepackage{booktabs}
\usepackage{relsize}
\usepackage{lineno}
\usepackage{comment}

\usepackage{braket}

\def\ubar{\overline{u}}

\def\lbar{\overline{\ell}}

\def\Babar{{\mbox{\slshape B\kern-0.1em{\smaller A}\kern-0.1em B\kern-0.1em{\smaller A\kern-0.2em R}}}}

\usepackage{cancel}

\usepackage{soul}

\usepackage[caption=false]{subfig}

\setstcolor{red}

\begin{document}

\title{Pinning down  $|\Delta c|=|\Delta u|=1$ couplings with rare charm baryon decays}

\author{Marcel Golz}
\email{marcel.golz@tu-dortmund.de}
\author{Gudrun Hiller}
\email{ghiller@physik.uni-dortmund.de}
\author{Tom Magorsch}
\email{tom.magorsch@tu-dortmund.de}
\affiliation{TU Dortmund University, Department of Physics, Otto-Hahn-Str.4, D-44221 Dortmund, Germany}

\begin{abstract}
We study the full angular distribution of  semileptonic rare charm baryon decays in which the secondary baryon undergoes weak decay
with sizable polarization parameter,  $\Xi_c^+\to\Sigma^+\,(\to p\pi^0)\ell^+\ell^-$, $\Xi_c^0\to\Lambda^0\,(\to p \pi^-)\ell^+\ell^-$ and  
$\Omega_c^0\to\Xi^0\,(\to \Lambda^{0} \pi^0)\ell^+\ell^-$.  Such self-analyzing decay chains allow for seven additional  observables compared to three-body decays such as $\Lambda_c \to p \ell^+ \ell^-$, with 
different sensitivities to the $|\Delta c|=|\Delta u|=1$ weak couplings.
Opportunities to test the standard model in $c \to u$ transitions with standard model null tests and other angular observables are worked out. 
We show that a joint model-independent analysis of  the leptonic $A_{\text{FB}}^\ell$, hadronic $A_{\text{FB}}^{\text{H}}$, and combined forward-backward asymmetry $A_{\text{FB}}^{\ell\text{H}}$ together with the fraction of longitudinally produced leptons, $F_L$, is able to pin down the dipole couplings $C_{7},C^\prime_{7}$ and the semileptonic (axial-) vector ones 
$C_{10},C^\prime_{9},C^\prime_{10}$.  $A_{\text{FB}}^{\text{H}}$ is also accessible with dineutrino  $c \to u \nu \bar \nu$  modes and probes right-handed currents.
\end{abstract}

\preprint{DO-TH 21/32}

\maketitle

\section{Introduction}

Flavor changing neutral currents of charm quarks are strongly suppressed  in the standard model (SM) by an efficient Glashow-Iliopoulos-Maiani (GIM) mechanism. 
At the same time sizable resonance contributions shadow new physics (NP) in simple observables such as branching ratios of  semileptonic  $c \to u \ell^+ \ell^-$-induced modes~\cite{Burdman:2001tf}.  This very GIM suppression, on the other hand,  along with approximate symmetries of the SM gives directions for clean  observables and null tests, which  probe a broad range of NP phenomena. Corresponding SM tests with  $|\Delta c|=|\Delta u|=1$ transitions complement  beyond standard model (BSM) searches with 
 strange and beauty quark processes and provide novel and unique insights into flavor from the up-quark sector. Several opportunities to test the SM have been worked out for 
 $D$-meson decays, {\textit{e.g.}}~\cite{Paul:2011ar,Cappiello:2012vg,Fajfer:2015mia,deBoer:2015boa,Gisbert:2020vjx,deBoer:2018buv, Bause:2019vpr}.

Rare semileptonic decays of charm baryons have been explored as a NP probe recently \cite{Meinel:2017ggx,Bause:2020xzj,Faisel:2020php,Golz:2021imq}.
In~\cite{Golz:2021imq}  we analyzed the NP sensitivity of  rare semileptonic decays of $\Lambda_c,\, \Xi_c$ and $\Omega_c$-baryons,  here collectively denoted by $B_0 \to B_1 \ell^+ \ell^-$ with  the initial (daughter) baryon denoted by $B_0 (B_1)$, see \cite{Bause:2020xzj} for  dineutrino modes  $B_0 \to B_1 \nu \bar \nu$.
In this work we consider (quasi-) four-body decays, where the $B_1$  further decays weakly to a hyperon $B_2$ and a pion.
Since kinematic observables, such as the direction of the $B_2$ momentum,  provide information on the spin of the decaying $B_1$ baryon,
these channels are termed 'self-analyzing'. Advantages of such modes for NP searches are well-known in  $b$-physics, notably using $\Lambda (1116) \to p \pi$ in rare decays of $\Lambda_b$-baryons, see for instance Ref.~\cite{Hiller:2001zj,Gutsche:2013pp,Boer:2014kda}.

In charm, we identify  the following decay channels suitable for polarization studies,\begin{align} \nonumber
\Xi_c^+\to\Sigma^+\,(\to p\pi^0)\ell^+\ell^-  \,  , & ~ \Xi_c^0\to\Lambda^0\,(\to p \pi^-)\ell^+\ell^-  \, , ~
\Omega_c^0\to\Xi^0\,(\to \Lambda^{0} \pi^0)\ell^+\ell^- \, , \\
\Xi_c^+\to\Sigma^+\,(\to p\pi^0) \nu \bar \nu   \,  , & ~ \Xi_c^0\to\Lambda^0\,(\to p \pi^-)  \nu \bar \nu   \, , ~
\Omega_c^0\to\Xi^0\,(\to \Lambda^{0} \pi^0)  \nu \bar \nu   \, , \nonumber
\end{align}
 since 
$\Sigma^+,\, \Lambda^0$ and $\Xi^0$ are self-analyzing, with sizable decay parameter  $\alpha$  (we do not consider $\Sigma^0 \to \Lambda \gamma$).
 The branching ratios and decay parameters of the secondary baryon decays are provided in Tab.~\ref{tab:list_channels}.
 Our aim is to work out  null tests and to complement   $|\Delta c|=|\Delta u|=1$  analyses of charmed meson decays, \textit{e.g.}~\cite{deBoer:2018buv, Bause:2019vpr}.  
 
 Requisite  vector and tensor form factors for $\Lambda_c\to p$ transitions have been computed on the  lattice~\cite{Meinel:2017ggx} and in quark models~\cite{Faustov:2018dkn}, and 
 for $\Xi_c\to\Sigma$  from Light cone sum rules~\cite{Azizi:2011mw}.
 As in~\cite{Golz:2021imq},  we employ $\Lambda_c\to p$ lattice form factors~\cite{Meinel:2017ggx} and relate them  to the  $\Xi_c,\,\Omega_c$ ones using  $SU(3)_F$-flavor symmetries, if applicable. This procedure is improvable with better knowledge of the form factors, however,  to explore  NP signals in SM null tests  a precise knowledge of form factors is not essential.
 
 None of the rare charm baryon modes has been observed so far, but the upper limit on the $\Lambda_c\to p\mu^+\mu^-$ branching ratio at $\sim 10^{-7}$ by LHCb~\cite{Aaij:2017nsd}
 is 
   close to the  estimated size of the resonance contributions~\cite{Golz:2021imq}. Limits on $\Lambda_c\to p e^+e^-$  and lepton flavor violating ones
$\Lambda_c\to p e^\pm \mu^\mp$   are at the level of  $\sim 10^{-5}$  by BaBar \cite{BaBar:2011ouc}.
 Semileptonic rare charm baryon decays are suitable for study at high luminosity flavor facilities, such as LHCb~\cite{Cerri:2018ypt}, Belle II~\cite{Kou:2018nap}, BES III~\cite{Ablikim:2019hff}, and possible future machines~\cite{Charm-TauFactory:2013cnj,Abada:2019lih}.

\begin{table}[!t]
 \centering
  \caption{Self-analyzing rare charm four-body decays $B_0 \to B_1 (\to B_2 \pi) \ell^+ \ell^-$ and information on the   branching ratios and weak decay parameters $\alpha$ of the secondary baryonic $B_1\to B_2 \pi$ decay~\cite{Zyla:2020zbs}. }
 \label{tab:list_channels}
 \begin{tabular}{c||c|c|c}
 & $\Xi_c^+\to\Sigma^+\,(\to p\pi^0)\ell^+\ell^-$ & $\Xi_c^0\to\Lambda^0\,(\to p \pi^-)\ell^+\ell^-$ & $\Omega_c^0\to\Xi^0\,(\to \Lambda^{0} \pi^0)\ell^+\ell^-$ \\
\hline 
$\mathcal{B}(B_1\to B_2 \pi)$ & $51.6\pm 0.3\%$ & $63.9\pm 0.5\%$ & $99.5\pm 0.0\%$ \\
$\alpha$ & $-0.98\pm 0.01$ & $0.73\pm 0.01$ & $-0.36\pm 0.01$ \\
 \end{tabular}
\end{table}

The plan of the paper is as follows: In Sec.~\ref{sec:theo} we discuss exclusive rare charm baryon decay modes within a low energy  effective field theory (EFT) framework, including
phenomenological resonance contributions.  We also present the full angular distribution for four-body baryon decays and review some of the simpler observables already
accessible with three-body decays.
We work out  the impact of the new null tests and other clean NP probes in Sec.~\ref{sec:nulltests}, and give an early stage strategy to disentangle NP Wilson coefficients. 
In Sec.~\ref{sec:further_nulltests} we present further null tests, based on more advanced angular observables, with decays into dineutrinos and for decays of polarized
charm baryons.
We conclude in Sec.~\ref{sec:conclusion}. We present the helicity amplitudes in terms of Wilson coefficients and form factors in App.~\ref{app:hel_amp}, and  provide details on the helicity amplitude description of the secondary weak decay in App.~\ref{app:hadr_hel}. In App.~\ref{app:angular_distr} we give the full angular distribution for initially polarized baryon decays.

\section{Theory of $|\Delta c|=|\Delta u|=1$ four-body baryon decays \label{sec:theo}}

We give general formulae for semileptonic rare charm baryon decays in the SM and beyond. In Sec.~\ref{Sec:heff} we introduce the weak Hamiltonian at the charm mass scale and discuss SM contributions. 
The fully differential distribution for the (quasi-)four-body decay of unpolarized charmed baryons is presented in Sec.~\ref{Sec:angular-dist}.

\subsection{An effective field theory approach to charm physics}\label{Sec:heff}

Consider the  weak effective Hamiltonian for $c\to u\ell^+\ell^-$ transitions
\begin{equation}
\mathcal{H}_{\rm eff} \supset -\frac{4G_F}{\sqrt2} \frac{\alpha_e}{4\pi}  \sum_{k=7,9,10} \bigl( C_k O_k + C_k^\prime O_k^\prime \bigr) \, ,
\label{eq:Heff}
\end{equation}
where $\alpha_e$ and $G_F$ denote the fine structure and Fermi's constant, respectively.
The dimension six operators are given as
\begin{equation}
\begin{split}
O_7 &= {m_c \over e} (\ubar_L \sigma_{\mu\nu} c_R) F^{\mu\nu} \,,\quad\quad \,O_7^\prime = {m_c \over e} (\ubar_R \sigma_{\mu\nu} c_L) F^{\mu\nu} \,, \\
O_9 &= (\ubar_L \gamma_\mu c_L) (\lbar \gamma^\mu \ell) \,, \quad\quad\quad \, O_9^\prime= (\ubar_R \gamma_\mu c_R) (\lbar \gamma^\mu \ell) \,, \\ 
O_{10} &= (\ubar_L \gamma_\mu c_L) (\lbar \gamma^\mu \gamma_5 \ell) \,,\quad\quad O_{10}^\prime= (\ubar_R \gamma_\mu c_R) (\lbar \gamma^\mu \gamma_5 \ell) \,,
\end{split}
\label{eq:operators}
\end{equation}
with  the electromagnetic field strength tensor $F^{\mu\nu}$, the chiral projectors $L=(1-\gamma_5)/2$, $R=(1+\gamma_5)/2$ and  $\sigma^{\mu\nu}=\frac{i}{2}\,[\gamma^\mu,\,\gamma^\nu]$.  For  the mass of the charm quark we use $m_c(m_c)=1.27\,\text{GeV}$, in the $\overline{\text{MS}}$ mass scheme.
SM contributions to the coefficients of the operators in Eq.~\eqref{eq:Heff} arise from four-quark operators at the $W$-mass scale and from intermediate resonances $M$, decaying electromagnetically to dileptons,
as in the quasi four-body decay chain  $B_0 \to B_1  M (\to \ell^+ \ell^-) \to B_1  \ell^+ \ell^-\to B_1 (\to B_2 \pi) \ell^+ \ell^-\to  B_2 \pi  \ell^+ \ell^- $. Note, the lifetime of the resonances $M=\omega, \rho, \phi$ is much shorter than the one of the  daughter baryons $B_1=\Sigma^+,\,\Lambda,\,\Xi^0$, which decay weakly after the dileptons have been produced. The resonance contributions are taken into account with a phenomenological ansatz, as
\begin{equation}
  \label{eq:resonances}
  C^{R}_{9}(q^{2}) = a_{\omega}e^{\text{i}\delta_{\omega}}\left(\frac{1}{q^{2} - m^{2}_{\omega} + \text{i}m_{\omega}\Gamma_{\omega}} - \frac{3}{q^{2} - m^{2}_{\rho} + \text{i}m_{\rho}\Gamma_{\rho}}\right)
        + \frac{a_{\phi}e^{\text{i}\delta_{\phi}}}{q^{2} - m^{2}_{\phi} + \text{i}m_{\phi}\Gamma_{\phi}} \, , 
\end{equation}
implying a contribution to   $O_9$.
Here, $m_M$ and $\Gamma_M$ denote the mass and total width of the meson $M$.
The strong phases $\delta_M$ are unknown and provide a significant amount of theoretical uncertainty.
We neglect effects from intermediate $\eta,\,\eta^\prime$ mesons as they are strongly localized and have a negligible effect on the (differential) branching ratio~\cite{Golz:2021imq}. 
We further use isospin to relate the  $\rho$ and $\omega$ contributions~\cite{Fajfer:2005ke}, as no data on any of  the $B_0 \to B_1 \rho$ branching ratios is available.
Experimental input on the parameters $a_M$ is provided in Tab.~\ref{tab:a_M}. 
Note that due to Belle's recent measurement of  
$\mathcal{B}(\Lambda_c^+\to p\omega)$~\cite{Belle:2021btl} the corresponding entry slightly differs from the one in~\cite{Golz:2021imq}.

 \begin{table}[!t]
 \centering
  \caption{Resonance parameters $a_{\omega},\,a_{\phi}$  defined in (\ref{eq:resonances})  for various rare charm baryon transitions, see text.}
 \label{tab:a_M}
 \begin{tabular}{l||c||c|c|c}
& $\Lambda_c\to p$ & $\Xi_c^+\to\Sigma^+$ & $\Xi_c^0\to\Lambda^0$ & $\Omega_c^0\to\Xi^0$ \\
\hline 
$a_\omega$ & $0.062 \pm 0.009$ & $\sim 0.06$ & $\sim 0.06$& $\sim 0.05$\\
$a_\phi$ & $0.110 \pm 0.008$ & $\sim0.1$  & $\sim 0.1$ & $\sim 0.09$ \\
 \end{tabular}
\end{table} 

For the form factors we use the same helicity-based definition as in~\cite{Golz:2021imq, Meinel:2017ggx}.
Form factors from lattice computations for $\Lambda_c\to p$ transitions are obtained in Ref.~\cite{Meinel:2017ggx}. We obtain the form factors for the baryon transitions studied in this work via flavor symmetries, see Refs.~\cite{Golz:2021imq, radiativeNico} for details. Consequently, we find for any of the ten form factors commonly denoted here as $f_{B_0 \to B_1}$
\begin{align}\label{eq:flavrel}
f_{\Lambda_c\to p}=f_{\Xi_c^+\to \Sigma^+}=\sqrt{6} f_{\Xi_c^0\to \Lambda^0} \simeq  f_{\Omega_c^0\to\Xi^0}.
\end{align}
We emphasize that all but the last relation follow from $SU(3)_F$ symmetry. The connection to the $\Omega_c$ is broken as it sits in a different multiplet. In absence of 
 form factor determinations for the latter at the same level as those for the $\Lambda_c \to p$ we use this simple relation to be able to make progress. We stress that this ansatz does not affect the null test features discussed in this work.
The relations (\ref{eq:flavrel})
have also been used for  $B_0 \to B_1 (\phi, \omega)$ to obtain the $a_M$ factors  for the  decays other than $\Lambda_c \to p  (\phi, \omega)$ presented in Tab.~\ref{tab:a_M}.
Specifically, the $\Lambda_c^+\to p$ parameters serve as an input to all other modes, as branching ratio data for the latter are not available. An exception is ${\mathcal{B}}(\Xi_c^0 \to \Lambda^0 \phi)=(4.9\pm1.5)\times10^{-4}$~\cite{Belle:2013ntc}, which gives  $a_\phi=0.080\pm0.013$, consistent  with the value in  Tab.~\ref{tab:a_M}.

Due to the severe GIM cancellation in rare charm decays, the perturbative SM contributions are overwhelmed by the effects from intermediate resonances: Perturbatively,  $C_7^{\text{eff}}(q^2)\sim 10^{-3}$, $C_9^{\text{eff}}(q^2)\sim 10^{-2}$, whereas the $\rho,\,\omega,\,\phi$ resonances yield $C_9^{R}(
q^2)\sim \mathcal{O}(10)$ on resonance peaks and $\sim \mathcal{O}(1)$ off peak, see~\cite{Golz:2021imq}, based on results in~\cite{deBoer:thesis, deBoer:2017way, deBoer:2015boa}. 
The primed Wilson coefficients  of (\ref{eq:operators})  are suppressed by $m_u/m_c$ and are negligible in the SM.
The Wilson coefficient $C_{10}$ vanishes in the SM, and therefore leptonic axialvector currents do, too, providing a prime opportunity for null test searches in charm. Electromagnetic loop contributions to the matrix element of 4-quark operators, or mixing,  induce contributions not exceeding permille level \cite{deBoer:2018buv}.

\subsection{Fully differential distribution  for $m_\ell \neq 0$ \label{Sec:angular-dist}}

We present the full differential decay distribution for four-body decays  $B_0 \to B_1 (\to B_2 \pi) \ell^+ \ell^-$.
A brief discussion of  $B_0 \to B_1 (\to B_2 \pi) \nu \bar \nu$ decays is deferred to  Sec.~\ref{sec:di}.
We compute the distribution using the   helicity formalism~\cite{Haber:1994pe, Gratrex:2015hna}  for unpolarized charmed baryons and keeping finite lepton masses, $m_\ell \neq 0$. Details on the  helicity amplitudes are given  in App.~\ref{app:hel_amp}. 
To be specific,  expressions are given for the decay $\Xi_c^+\to \Sigma^+(\to p \pi^0)\ell^+\ell^-$, however, with replacements of masses, form factors and $B_1 \to B_2 \pi$ branching ratios, the same holds for any of the other modes in Tab.~\ref{tab:list_channels}.
The fully differential distribution can be parameterized in terms of the ten $q^2$-dependent angular observables $K_i =K_i(q^2)$ as
\begin{equation}
\begin{split}
  \frac{\text{d}^4\Gamma}{\text{d}q^2\text{d}\cos\theta_\ell \text{d}\cos\theta_\pi \text{d}\phi}=\frac{3}{8\pi}\cdot\bigg[&\,\,K_{1ss}\,\sin^2\theta_\ell\,+\,K_{1cc}\,\cos^2\theta_\ell\,+\,K_{1c}\,\cos\theta_\ell\\
  +&\left( K_{2ss}\,\sin^2\theta_\ell\,+\,K_{2cc}\,\cos^2\theta_\ell\,+\,K_{2c}\,\cos\theta_\ell\right)\cos\theta_\pi \\
  +&\left( K_{3sc}\,\sin\theta_\ell\cos\theta_\ell + K_{3s}\sin\theta_\ell\right)\sin\theta_\pi\sin\phi \\
  +&\left( K_{4sc}\,\sin\theta_\ell\cos\theta_\ell + K_{4s}\sin\theta_\ell\right)\sin\theta_\pi\cos\phi\bigg]\,.
  \end{split}
  \label{eq:angl_distr}
\end{equation}
Here,  $\theta_\ell$ is the angle of the $\ell^+$ with respect to the negative direction of flight of the charmed baryon ($\Xi^+_c$) in the dilepton rest frame. Similarly, $\theta_\pi$ is the angle between the momentum of the final state ($B_2$)baryon ($p$) and the negative direction of flight of the $B_1$ baryon ($\Sigma^+$) in the proton-pion center-of-mass frame. The azimuthal angle $\phi$ describes the angle between the dilepton and the $p\pi^0$ decay planes. 
The allowed regions for the angles $\theta_\ell$, $\theta_\pi$, $\phi$ are  $-1 \leq \cos\theta_\ell \leq +1$, $-1<\cos\theta_\pi<1$ and $0<\phi<2\pi$.

The $q^2$-dependent coefficients $K_i$ are given as~\cite{Gutsche:2013pp}\footnote{We adapt the notation of helicity  expressions $I^{mm^{\prime}}_{iP}, i=1,2,3,4$ from~\cite{Gutsche:2013pp}, however use them to formulate angular observables in a notation similar to~\cite{Boer:2014kda}. Note that we dropped the subscript $P$ from $I^{mm^{\prime}}_{2},\,I^{mm^{\prime}}_{3}$ since these two interference terms are parity-even.}
\begin{align}
  \label{eq:2}
  \begin{split}
    \frac{K_{1ss}}{\mathcal{B}(\Sigma^+\to p \pi^0)} &=\phantom{-2}q^{2}v^{2}\left(\frac{1}{2}U^{11+22} + L^{11+22}\right) + 4m^{2}_{\ell}\left(U^{11}+L^{11}+S^{22}\right),\\
    \frac{K_{1cc}}{\mathcal{B}(\Sigma^+\to p \pi^0)} &=\phantom{-2}q^{2}v^{2}U^{11+22} + 4m^{2}_{\ell}\left(U^{11}+L^{11}+S^{22}\right),\\
    \frac{K_{1c}}{\mathcal{B}(\Sigma^+\to p \pi^0)} &=-2q^{2}vP^{12},\\
    \frac{K_{2ss}}{\mathcal{B}(\Sigma^+\to p \pi^0)\cdot \alpha} &=\phantom{-2} q^{2}v^{2}\left(\frac{1}{2}P^{11+22} + L_P^{11+22}\right) + 4m^{2}_{\ell}\left(P^{11}+L_P^{11}+S_P^{22}\right),\\
    \frac{K_{2cc}}{\mathcal{B}(\Sigma^+\to p \pi^0)\cdot \alpha} &=\phantom{-2} q^{2}v^{2}P^{11+22} + 4m^{2}_{\ell}\left(P^{11}+L_P^{11}+S_P^{22}\right),\\
    \frac{K_{2c}}{\mathcal{B}(\Sigma^+\to p \pi^0)\cdot \alpha} &=-2q^{2}vU^{12},\\
    \frac{K_{3sc}}{\mathcal{B}(\Sigma^+\to p \pi^0)\cdot \alpha} &={-2\sqrt{2}q^{2}v^{2}I_{2}^{11+22},}\\
    \frac{K_{3s}}{\mathcal{B}(\Sigma^+\to p \pi^0)\cdot \alpha} &={\phantom{-}4\sqrt{2}q^{2}v I_{4P}^{12},}\\
    \frac{K_{4sc}}{\mathcal{B}(\Sigma^+\to p \pi^0)\cdot \alpha} &={\phantom{-}2\sqrt{2}q^{2}v^{2}I_{1P}^{11+22},}\\
    \frac{K_{4s}}{\mathcal{B}(\Sigma^+\to p \pi^0)\cdot \alpha} &={-4\sqrt{2}q^{2}v I_{3}^{12},}\\
  \end{split}
\end{align}
in agreement with our own computation and~\cite{Blake:2017une}. Here,
$v = \sqrt{1-\frac{4m^{2}_{\ell}}{q^{2}}}$, $U^{11+22}=U^{11} + U^{22}$ and likewise for $L,\,P,\,I_{1P},\,I_{2}$. The $q^2$-dependent  terms $U,\, L,\, S,\, P,\,L_P,\,S_P$ denote quadratic expressions of helicity amplitudes and correspond to unpolarized transverse, longitudinal, scalar, transverse parity-odd, longitudinal parity-odd and scalar parity-odd contributions, respectively. The coefficients $I_{1P},\,I_{4P}$ and $I_{2},\,I_{3}$ correspond to longitudinal-transverse interference terms, where the subscript $P$ refers to  the parity-odd ones. We refer  to App.~\ref{app:hel_amp} for  expressions in terms of Wilson coefficients and hadronic form factors, $f_i(q^2), g_i(q^2), i=+,\perp,0$ and $h_j(q^2), \tilde h_j(q^2), j=+, \perp$, which are defined in Ref.~\cite{Meinel:2017ggx, Golz:2021imq}.

The GIM mechanism is responsible for the absence of leptonic axial-vector currents in rare charm decays. Therefore, neglecting higher order electromagnetic contributions to 
$C_{10}$~\cite{deBoer:2018buv},
\begin{align}K_{1c}^{\rm SM} =K_{2c}^{\rm SM}=K_{3s}^{\rm SM}=K_{4s}^{\rm SM}=0 \, .
\label{eq:null}
\end{align}
 At the same time, these angular observables serve as clean null tests of the SM.
The first one, $K_{1c}$, has already been studied  in $\Lambda_c \to p \mu^+ \mu^-$~\cite{Meinel:2017ggx}  and  three-body $1/2 \to 1/2  \ell^+ \ell^-$ decays of 
$\Lambda_c,\Xi_c$ and $\Omega_c$'s~\cite{Golz:2021imq}.
The other three null tests, $K_{2c}, K_{3s}$ and $K_{4s}$  are a new result of this work. They become accessible in four-body decays, and vanish for $\alpha=0$.
We analyze the NP sensitivity in Section \ref{sec:nulltests}.

Let us recap basic features of the distribution Eq.~\eqref{eq:angl_distr}. If both  $\theta_\pi$ and $\phi$ are not measured, only the first line survives and  one recovers the double differential distribution for three-body decays:
\begin{equation}
  \frac{\text{d}^2\Gamma}{\text{d}q^2\text{d}\cos\theta_\ell}=\int_{-1}^{1} \, \! \!\int_{0}^{2\pi} \! \! \! \frac{\text{d}^4\Gamma}{\text{d}q^2 \text{d}\cos\theta_\ell \text{d} \! \cos\theta_\pi\text{d} \phi} \,\text{d} \phi \text{d} \! \cos\theta_\pi= \frac{3}{2}\,(K_{1ss}\,\sin^2\theta_\ell\,+\,K_{1cc}\,\cos^2\theta_\ell\,+\,K_{1c}\,\cos\theta_\ell)\,.
  \label{eq:angl_distr-3}
\end{equation}
From here follows the 
$q^2$-differential decay rate  
\begin{equation}
\frac{\text{d}\Gamma}{\text{d}q^2}=\int_{-1}^{1}     \frac{\text{d}^2\Gamma}{\text{d}q^2\text{d}\cos\theta_\ell}      \text{d} \! \cos\theta_\ell=2\,K_{1ss}+K_{1cc}\, ,
\label{eq:diffbrratio}
\end{equation}
the  longitudinal fraction of the dilepton system, $F_L$,
\begin{equation}
 F_L=\frac{2\,K_{1ss}-K_{1cc}}{2\,K_{1ss}+K_{1cc}}\,, 
 \label{eq:fl}
\end{equation}
and the forward-backward asymmetry of the leptonic scattering angle, $A_{\text{FB}}^{\ell}$,
\begin{equation}
\begin{split}\label{eq:afb_l}
A_{\text{FB}}^{\ell}&=\frac{1}{\text{d}\Gamma /\text{d}q^2}\,\left[\int_0^1\,-\,\int_{-1}^0\right]          \frac{\text{d}^2\Gamma}{\text{d}q^2\text{d}\cos\theta_\ell}      \text{d} \! \cos\theta_\ell
= \frac{3}{2}\,\frac{K_{1c}}{2\,K_{1ss}+K_{1cc}}\,,
\end{split}
\end{equation}
see~\cite{Golz:2021imq} for a detailed discussion of the phenomenology in and beyond the SM.
 
 Kinematic endpoints are $q^2_{\text{min}}=4 m_\ell^2$,  corresponding to maximum hadronic recoil, and 
 $q_{\text{max}}^2=(m_{B_0} - m_{B_1})^2$,
 corresponding to zero hadronic recoil. The latter is subject to symmetry relations,  enforcing 
 $K_{1ss}=K_{1cc}$, hence $F_L=1/3$  and similarly $K_{2ss}=K_{2cc}$ model-independently at this point  \cite{Hiller:2021zth}. 
 These relations hold  also at  the other end of the spectrum, at $q^2_{\text{min}}$, because here  the four-momenta of the leptons coincide which leads also to a reduction of Lorentz structures~\cite{Golz:2021imq}.

The integrated decay rate is obtained as 
\begin{align}
\Gamma=\int_{q^2_{\text{min}}}^{q^2_{\text{max}}}\,(2\,K_{1ss}+K_{1cc})\,\text{d}q^2 \, , 
\end{align}
where phase space cuts may be applied. Integrating the full $q^2$ region with $\pm 40\,$MeV cuts~\cite{Aaij:2017nsd} around  the $\omega$ and the $\phi$ resonances, we find 
\begin{equation} \label{eq:Brs}
\begin{split}
\mathcal{B}(\Xi_c^+\to \Sigma^+(\to p\pi^0)\mu^+\mu^-) \sim 1.8\times10^{-8}\,,\\
\mathcal{B}(\Xi_c^0\to \Lambda^0(\to p\pi^-)\mu^+\mu^-) \sim 2.4\times10^{-9}\,,\\
\mathcal{B}(\Omega_c^0\to \Xi^0(\to \Lambda^0\pi^0)\mu^+\mu^-) \sim 2.5\times10^{-8}\,,
\end{split}
\end{equation}
in agreement with results in Ref.~\cite{Golz:2021imq} multiplied
with $\mathcal{B}(B_1\to B_2 \pi)$ given in Tab.~\ref{tab:list_channels}.

\section{Probing NP with  Angular Observables} \label{sec:nulltests}

In this section we discuss angular observables in rare, semileptonic charm baryon decays that can  cleanly signal NP, and work out sensitivities to specific Wilson coefficients.
The full angular distribution (\ref{eq:angl_distr})  features four GIM-based null tests, $K_{1c}, K_{2c}, K_{3s}$ and $K_{4s}$, which vanish  in the SM (\ref{eq:null}).
In addition,  $F_L$  (\ref{eq:fl}) is also a sensitive probe of NP~\cite{Golz:2021imq}. In Sec.~\ref{sec:simple} we work out BSM signatures in $F_L$ and in a similarly simple and sensitive observable, the hadronic  forward-backward asymmetry, $\sim 2K_{2ss}+K_{2cc}$.
$K_{1c}$ and $K_{2c}$ correspond to the leptonic and combined leptonic-hadronic forward-backward asymmetries, respectively. They are discussed in Sec.~\ref{sec:afbnull}.
 Null tests in the longitudinal-transverse interference terms $K_{4s} \sim I_{3}^{12}$ and $K_{3s} \sim I_{4P}^{12}\,$ are analyzed in the next Sec.~\ref{sec:further_nulltests}.
We summarize a strategy to disentangle Wilson coefficients based on three asymmetries and $F_L$ in Sec.~\ref{sec:final_analysis}.

\subsection{$A_{\text{FB}}^{\text{H}}$ and $F_L$  \label{sec:simple}}

The hadronic forward-backward asymmetry $A_{\text{FB}}^{\text{H}}$ is defined similar to  the leptonic one, $A_{\text{FB}}^{\ell}$, Eq.~\eqref{eq:afb_l}, as
\begin{equation}
\begin{split}\label{eq:afb_H}
A_{\text{FB}}^{\text{H}}&=\frac{1}{\text{d}\Gamma /\text{d}q^2}\,\left[\int_0^1\,-\,\int_{-1}^0\right]\int_{-1}^1\int_0^{2\pi}\frac{\text{d}^4\Gamma}{\text{d}q^2 \text{d}\cos\theta_\ell\text{d}\cos\theta_\pi\text{d}\phi} \text{d} \phi \text{d}\cos \theta_\ell \text{d}\cos \theta_\pi \\ &= \frac{1}{2}\,\frac{2\,K_{2ss}+K_{2cc}}{2\,K_{1ss}+K_{1cc}}\,.
\end{split}
\end{equation}
Unlike  $A_{\text{FB}}^{\ell}$, $A_{\text{FB}}^{\text{H}}$  is not a null test of the SM, however, it turns out to be a highly  sensitive probe of right-handed quark currents as illustrated in the left plot of Fig.~\ref{fig:afb_H}. 
Here, the orange curve displays the SM expectation, and several NP benchmarks are shown in red, green and blue. The brackets in the subscripts are understood as  \textit{or}, for instance, $C_{9,\,(10)}^\prime=0.5$ is short  for $C_9^\prime=0.5$ \textit{or} $C_{10}^\prime=0.5$. 
We learn that $A_{\text{FB}}^{\text{H}}$  is sensitive to $C_7^\prime$, $C_9^\prime$ and $C_{10}^\prime$.

$A_{\text{FB}}^{\text{H}}$ shares features with $F_L$ (\ref{eq:fl}), shown in the right plot of Fig.~\ref{fig:afb_H}: Cancellation of hadronic uncertainties in the SM (orange), strong sensitivity to NP contributions in \textit{some} Wilson coefficients and large uncertainties in NP scenarios due to unknown strong phases, observed previously for 
$F_L$ in~\cite{Golz:2021imq}. The main differences between these two angular observables are the following:

\begin{figure}[h!]\centering
\includegraphics[width=0.5\textwidth]{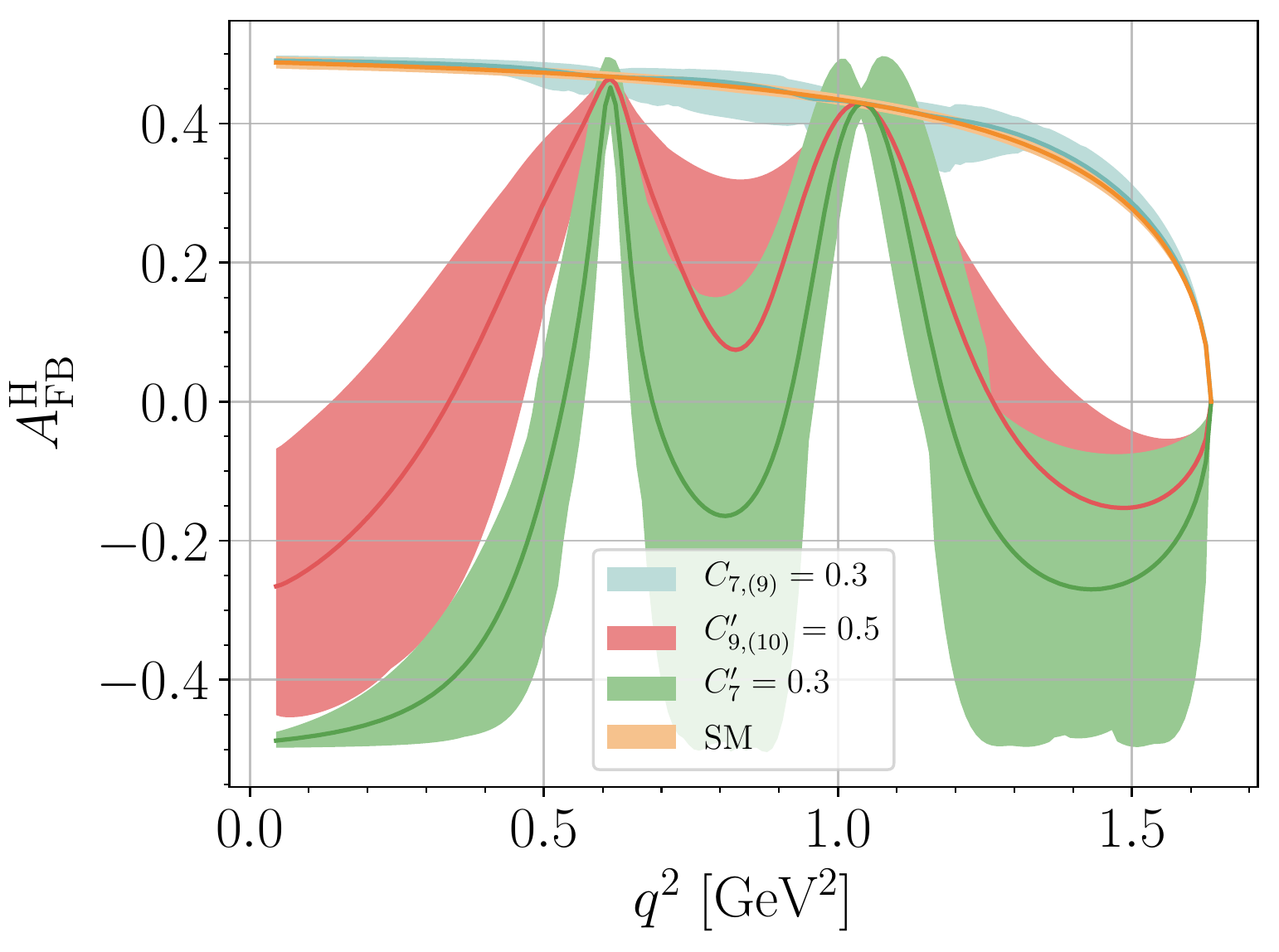}
\includegraphics[width=0.48\textwidth]{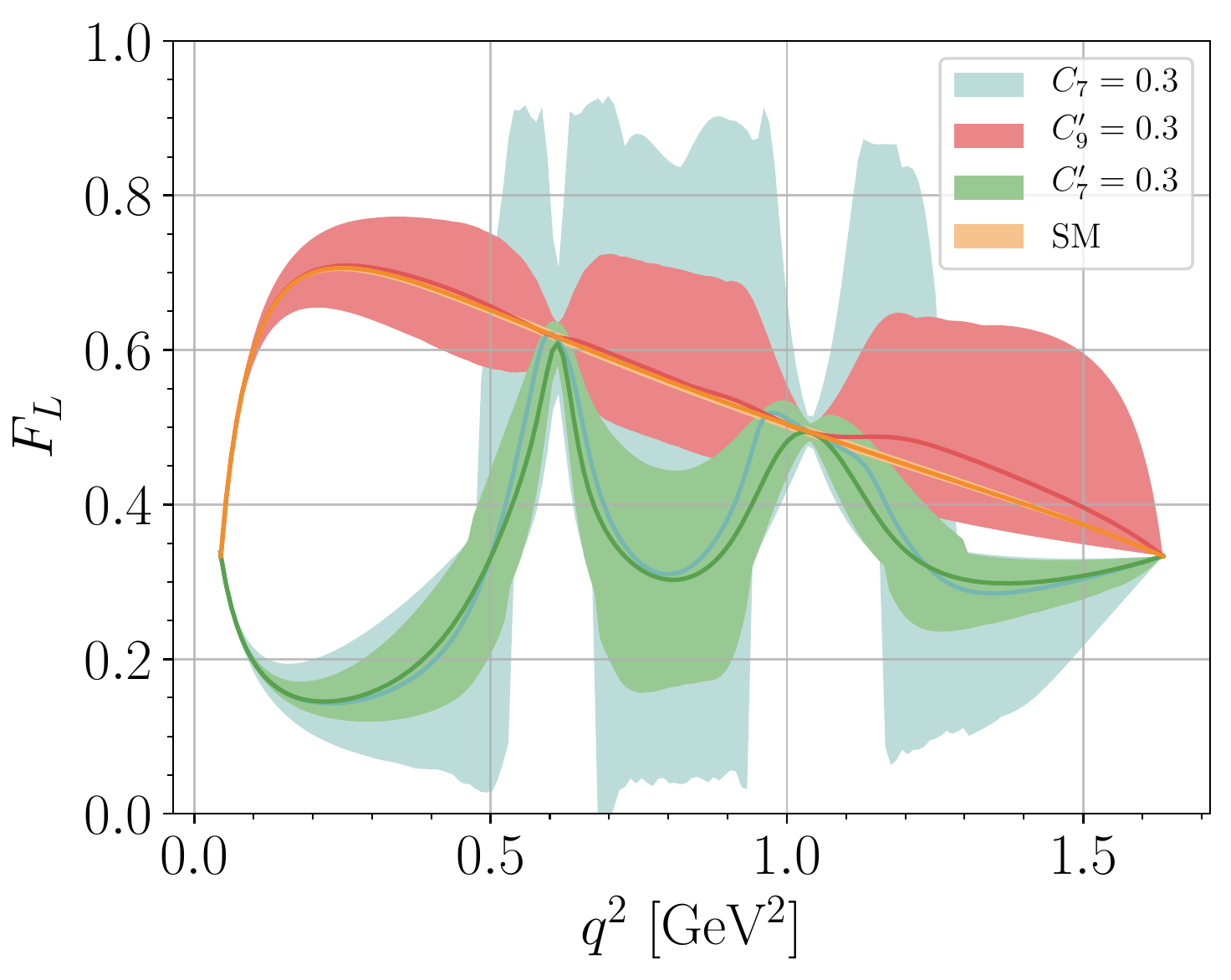}
\caption{The hadronic forward-backward asymmetry $A_{\text{FB}}^{\text{H}}$ (\ref{eq:afb_H}) (left plot) for $\Xi^+_c\to\Sigma^+(\to p\pi^0)\mu^+\mu^-$ decays in the SM  (orange) and in NP scenarios with $C_{7}$ or $C_{9}=0.3$, $C_{9}^\prime$ or $C_{10}^\prime=0.5$ and $C_7^\prime=0.3$ in blue, red and green, respectively. A NP scenario with $C_{10}$ only is not shown, as it is indistinguishable from the SM. The right plot  shows the fraction of longitudinally polarized dimuons $F_L$ (\ref{eq:fl}) in the SM (orange) and NP scenarios $C_7=0.3$, $C_9^\prime=0.5$ and $C_7^\prime=0.3$ in blue, red and green, respectively. Scenarios with $C_9$ and $C_{10}^{(\prime)}$ can not be distinguished from the SM with $F_L$ and are not shown. The width of the bands stem predominantly from unknown strong phases.}
\label{fig:afb_H}
\end{figure}

\begin{itemize}
  \item $F_L=1/3$ at both kinematic endpoints of  maximum and zero recoil, whereas $A_{\text{FB}}^{\text{H}}$ is unconstrained at low $q^2$ and vanishes at maximum $q^2$.
  \item $F_L$ is mostly sensitive to radiative dipole couplings $C_7$  and $C_7^\prime$ , see the blue and green bands in the right  plot of Fig.~\ref{fig:afb_H}. $A_{\text{FB}}^{\text{H}}$ is similar (equal) to the SM in scenarios involving $C_{7}$ or $C_{9}$ ($C_{10}$), but strongly altered in scenarios involving right-handed currents $C^\prime_7,\,C^\prime_9,\,C_{10}^\prime$, see the green and red bands in the left plot of Fig.~\ref{fig:afb_H}.
\end{itemize}

  The different impact  of left-handed and right-handed NP contributions to  $A^{\text{H}}_{\text{FB}}$ can be attributed to the parity behavior of the angular observables. While $K_{1ss}$ and $K_{1cc}$ are P-even observables, $K_{2ss}$ and $K_{2cc}$ are P-odd. This leads to cancellations between numerator and denominator only in the case of left-handed contributions. To illustrate this consider $A^{\text{H}}_{\text{FB}}$ for  $m_{\ell}=0$ in  scenarios with $C_{9}$ and 
  $C^{(\prime)}_{10}$ 
  and all other NP coefficients switched off.
  It can be written as 
  \begin{equation}
    \label{eq:4}
    A^{\text{H}}_{\text{FB}}=-\alpha\cdot\frac{\left(\left|C_{9}\right|^{2}+ \left|C_{10}\right|^{2}- \left|C^{\prime}_{10}\right|^{2}\right)A(q^{2})\sqrt{s_{+}s_{-}}}{ 
\left(    \left( \left|C_{9}\right|^{2}+ \left|C_{10}-C_{10}^\prime \right|^{2} \ \right)
    B(q^{2})s_{+} + 
    \left( \left|C_{9}\right|^{2}+ \left|C_{10}+C_{10}^\prime \right|^{2} \right)
    C(q^{2})s_{-}\right)} \, ,   \end{equation}
where $A(q^{2}),B(q^{2})$ and $C(q^{2})$ 
contain form factors and kinematics and are given in App.~\ref{app:hel_amp}, and $s_{\pm} = (m_{B_0} \pm m_{B_1})^2 - q^2$.
For $C^{(\prime)}_{10}=0$ the coefficient $C_{9}$ cancels as in $F_{L}$, leading to the thin SM (orange) band. 
For $C_{10}\neq 0$ the same effect happens and  $\left|C_{9}\right|^{2}+\left|C_{10}\right|^{2}$ drops out. On the other hand,  for $C^{\prime}_{10}\neq 0$ the numerator is proportional to  $\left|C_{9}\right|^{2}-\left|C^{\prime}_{10}\right|^{2}$, which does not cancel against the denominator and leads to NP deviations with  $q^{2}$-shape driven by  $C^{R}_{9}(q^{2})$.
The same arguments holds for dipole couplings $C_{7}$ and  $C^{(\prime)}_{7}$:   $A^{\text{H}}_{\text{FB}}$ is strongly sensitive to the latter, but not the former.
 Note,  interference terms between $C_{9}$ and $C_{7}$ softly break the exact cancellation (blue band around the SM). Again we stress that the requisite additional minus sign in front of  the primed Wilson coefficients arises because $K_{2ss}$ and $K_{2cc}$ are P-odd.

\subsection{$A_{\text{FB}}^{\ell\text{H}}$ and $A_{\text{FB}}^{\ell}$  \label{sec:afbnull}}

A third forward-backward asymmetry arises from combining leptonic and hadronic ones, $A_{\text{FB}}^{\ell\text{H}}$,
\begin{equation}
\begin{split}\label{eq:afb_lH}
A_{\text{FB}}^{\ell\text{H}}&=\frac{1}{\text{d}\Gamma /\text{d}q^2}\,\left[\int_0^1\,-\,\int_{-1}^0\right]\left[\int_0^1\,-\,\int_{-1}^0\right]\int_0^{2\pi}\frac{\text{d}^4\Gamma}{\text{d}q^2\text{d}\cos\theta_\ell\text{d}\cos\theta_\pi}  \text{d}\phi\text{d}\cos \theta_\ell\text{d}\cos \theta_\pi \\ &= \frac{3}{4}\,\frac{K_{2c}}{2\,K_{1ss}+K_{1cc}}\,.
\end{split}
\end{equation}
It is yet another charming null test  of the SM, because $C_{10}$ or $C_{10}^\prime$  are required to observe a non-vanishing signal.

\begin{figure}[h!]\centering
  \includegraphics[width=0.49\textwidth]{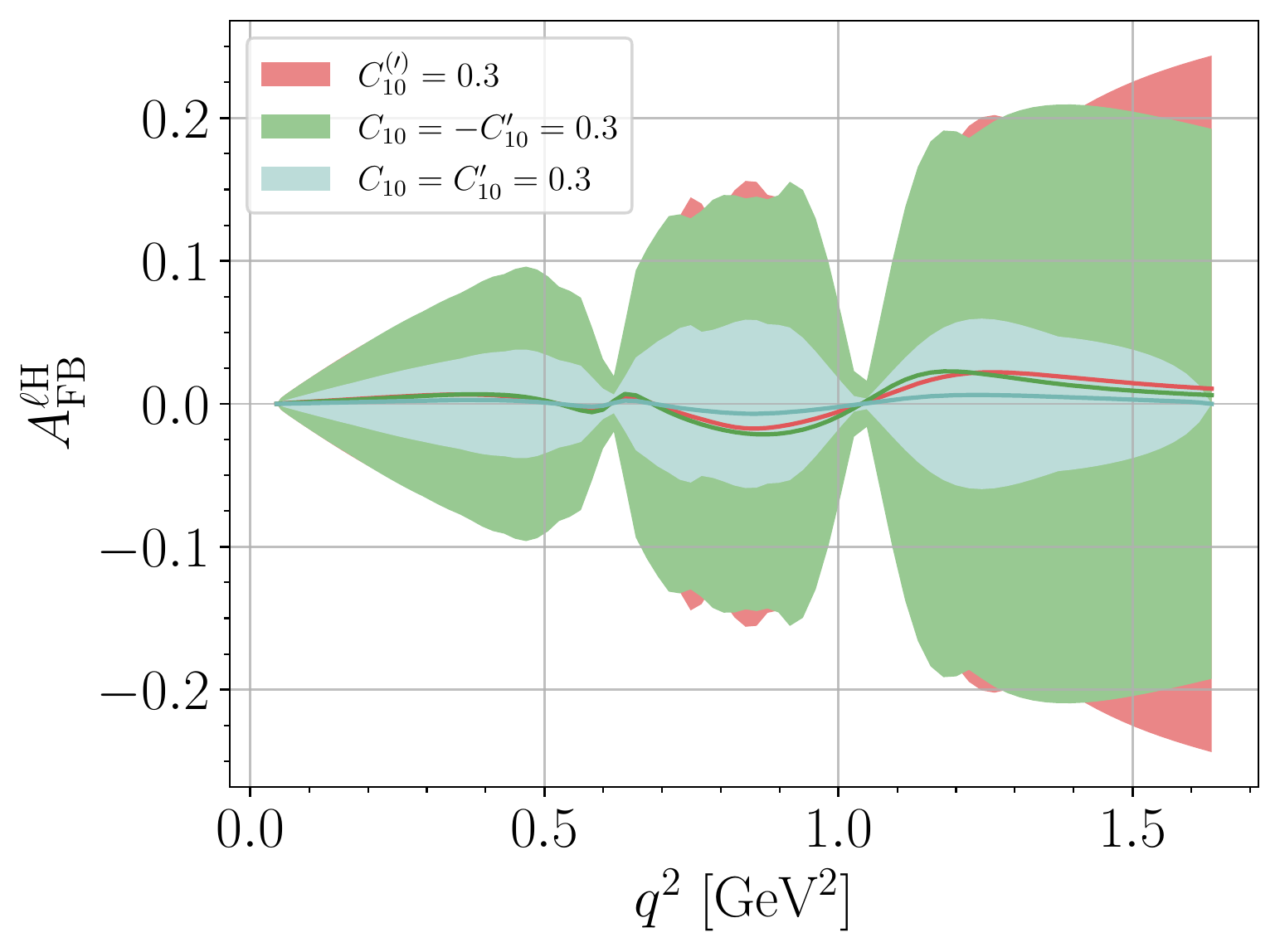}
  \includegraphics[width=0.49\textwidth]{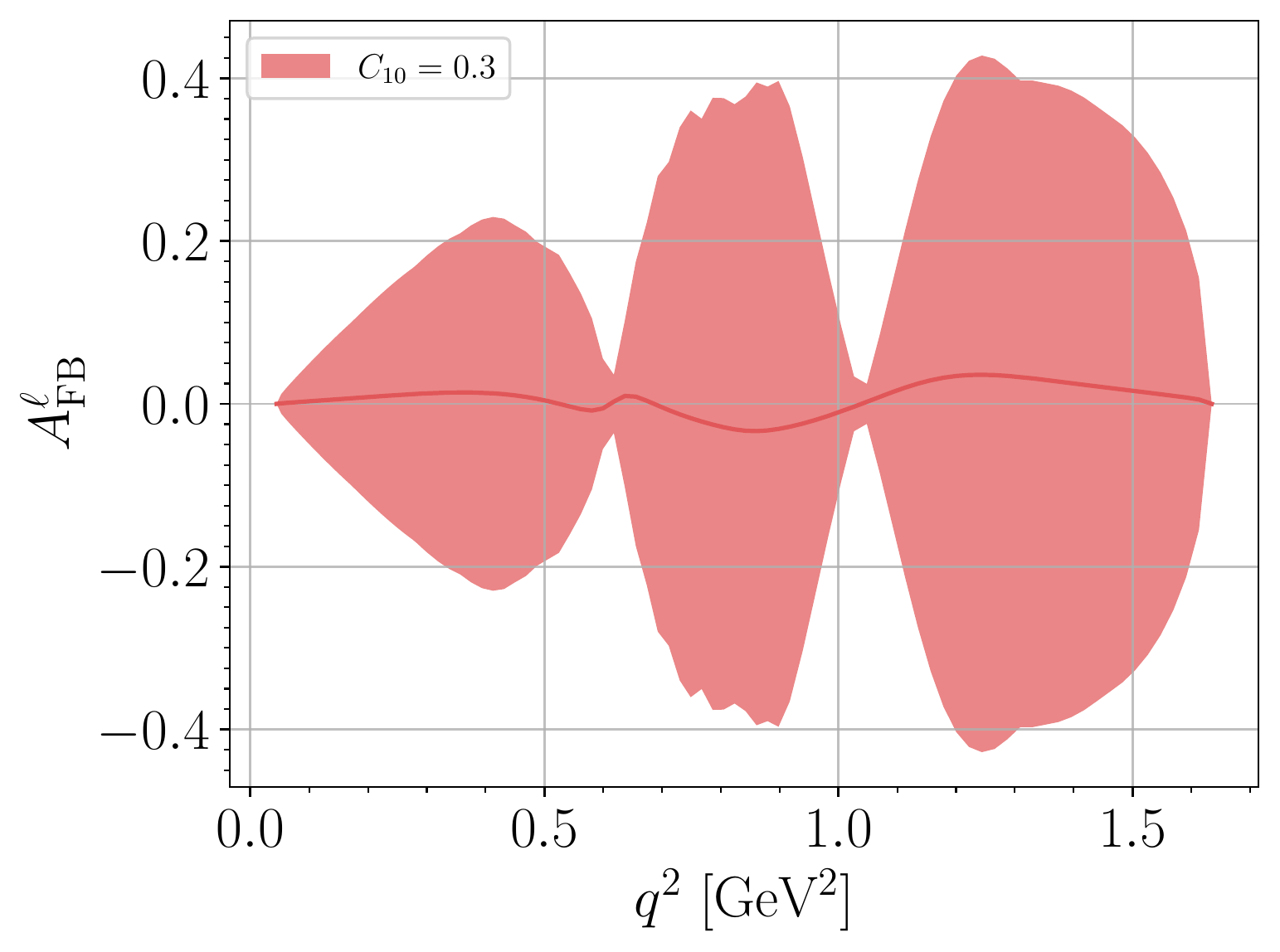}
\caption{The left plot shows the  forward-backward asymmetry in both hadronic and leptonic scattering angles, $A_{\text{FB}}^{\ell\text{H}}$ (\ref{eq:afb_lH})  for  $\Xi^+_c\to\Sigma^+(\to p\pi^0)\mu^+\mu^-$  decays in NP scenarios with $C_{10}$ or $C_{10}^\prime=0.3$, $C_{10}=-C_{10}^\prime=0.3$ and $C_{10}=C_{10}^\prime=0.3$ in red, green and blue, respectively. The right plot shows $A_{\text{FB}}^{\ell}$ (\ref{eq:afb_l})  only in a $C_{10}=0.3$ scenario. The width of the bands stem predominantly from unknown strong phases.
Both $A^{\ell \text{H}}_{\text{FB}}$ and $A^{\ell}_{\text{FB}}$ are null tests of the SM (\ref{eq:null}).}
\label{fig:afb_lH}
\end{figure}

In Fig.~\ref{fig:afb_lH} $A_{\text{FB}}^{\ell\text{H}}$ is shown (left plot)  for  three different NP scenarios in red, green and blue for $C_{10}$ or $C_{10}^\prime=0.3$, $C_{10}=-C_{10}^\prime=0.3$ and $C_{10}=C_{10}^\prime=0.3$, respectively. 
These benchmarks  are chosen to illustrate the following: 
 Firstly, $C_{10}$ and $C_{10}^\prime$ contributions are indistinguishable within the large uncertainties induced by unknown strong phases entering in Eq.~\eqref{eq:resonances} and varied in the plot. Secondly, 
 a scenario with $C_{10}=C_{10}^\prime$ leads to a partial cancellation of contributions leading to a decreased signal with respect to $C_{10}=-C_{10}^\prime$ scenarios. 
We also show $A_{\text{FB}}^{\ell}$ (\ref{eq:afb_l}) in Fig.~\ref{fig:afb_lH} (right plot)  for  $C_{10}=0.3$. We recall that 
$A_{\text{FB}}^{\ell}$ is a charm specific null test with sensitivity to the axial-vector coupling $C_{10}$ down to the percent level. 
 $A_{\text{FB}}^{\ell}$ vanishes at both, the low and the high $q^{2}$ endpoints.

The main benefit in  studying $A_{\text{FB}}^{\ell\text{H}}$ in addition to $A_{\text{FB}}^{\ell}$ is complementarity. As pointed out in Ref.~\cite{Golz:2021imq}, $A_{\text{FB}}^{\ell}$ has sensitivity to $C_{10}$, but not necessarily $C_{10}^\prime$, as this would require also  NP contributions in $C_{9}^\prime$. In $A_{\text{FB}}^{\ell\text{H}}$ interference terms of 
type $C_9C_{10}^\prime$ exist, which are needed to observe a NP signal in a
$C_{10}^\prime$-only scenario.
In addition, $A^{\ell\text{H}}_{\text{FB}}$ does not necessarily vanish at the high $q^{2}$ endpoint, and rather assumes a model-dependent value  \cite{Hiller:2021zth}.

\subsection{Model-independent analysis}\label{sec:final_analysis}

To outline the strategy for disentangling NP contributions in charm baryon decays, we first summarize the sensitivities to single Wilson coefficients. 
The observables $A^{\ell}_{\text{FB}}$ and $F_L$ appear  in three-body decays, while  $A^{\text{H}}_{\text{FB}}$  and $A^{\ell\text{H}}_{\text{FB}}$ are
arise in self-analyzing  four-body decays discussed in this work.
\begin{enumerate}
  \item $A^{\ell}_{\text{FB}}$ is a null test that probes $C_{10}$
  \item $A^{\ell\text{H}}_{\text{FB}}$ is a null test that probes $C_{10}$ and $C^{\prime}_{10}$
  \item $F_{L}$ probes $C_{7}$ and $C^{\prime}_{7}$
  \item $A^{\text{H}}_{\text{FB}}$ probes $C^{\prime}_{7}$, $C^{\prime}_{9}$ and $C^{\prime}_{10}$
\end{enumerate}
If there is no signal observed for the null tests $A^{\ell}_{\text{FB}}$ and $A^{\ell\text{H}}_{\text{FB}}$,  one concludes that both $C_{10}$ and $C^{\prime}_{10}$ are well below the percent level. In a next step $F_{L}$ can be used to 
probe dipole contributions $C_{7}$ or $C^{\prime}_{7}$. To differentiate between the left-handed and right-handed dipole operators, $A^{\text{H}}_{\text{FB}}$ can be employed. Similarly $C^{\prime}_{9}$ can be extracted from $A^{\text{H}}_{\text{FB}}$, if $F_{L}$ is SM-like. In a scenario with non-vanishing null tests, NP contributions to $C_{10}^{(\prime)}$ are evident. Here, again $A^{\text{H}}_{\text{FB}}$ differentiates $C_{10}$ and $C_{10}^\prime$. In addition, $A^{\ell}_{\text{FB}}$ and $A^{\ell\text{H}}_{\text{FB}}$ reveal information on $C_{10}^\prime$ contributions.
The only coefficient which can not be probed efficiently is $C_{9}$, as it is dominated by the resonances $C^{R}_{9}$. As anticipated already in~\cite{Golz:2021imq}, a future simultaneous fit of Wilson coefficients and resonance parameters is then needed.

\section{Further null tests}\label{sec:further_nulltests}

In this section we discuss further null test opportunities for  rare  charm baryon decays.
We begin with the angular observables $K_{3s}$ and $K_{4s}$ (\ref{eq:null}) in Sec.~\ref{sec:3s4s},  discuss dineutrino modes in Sec.~\ref{sec:di} and present null tests that become available if the initial charm baryon is polarized (Sec.~\ref{sec:pol}).
The study of charged lepton flavor violating  modes offers even more clean null tests but is beyond the scope of this work.

\subsection{$K_{3s}$ and $K_{4s}$  \label{sec:3s4s}}

The angular observables $K_{3s}$ and $K_{4s}$ vanish in the SM (\ref{eq:null}) or any SM extension with vanishing $C_{10}$ and $C^{\prime}_{10}$.
Both  $K_{3s}$ and $K_{4s}$ contain terms of the form $C_{9}C_{10}$ and $C_{9}C^{\prime}_{10}$, just like $K_{2c} \propto A^{\ell\text{H}}_{\text{FB}}$,
and unlike $K_{1c} \propto A^{\ell}_{\text{FB}}$. The latter requires additional NP coefficients to be sensitive to $C_{10}^\prime$.
This way, $K_{3s}$ and $K_{4s}$ are structurally similar to $A^{\ell\text{H}}_{\text{FB}}$, discussed in Sec.~\ref{sec:afbnull}.
All three of them probe $C_{10}$ and $C_{10}^\prime$, although with different combinations of form factors and NP coefficients.

\begin{figure}[h!]\centering
  \includegraphics[width=0.48\textwidth]{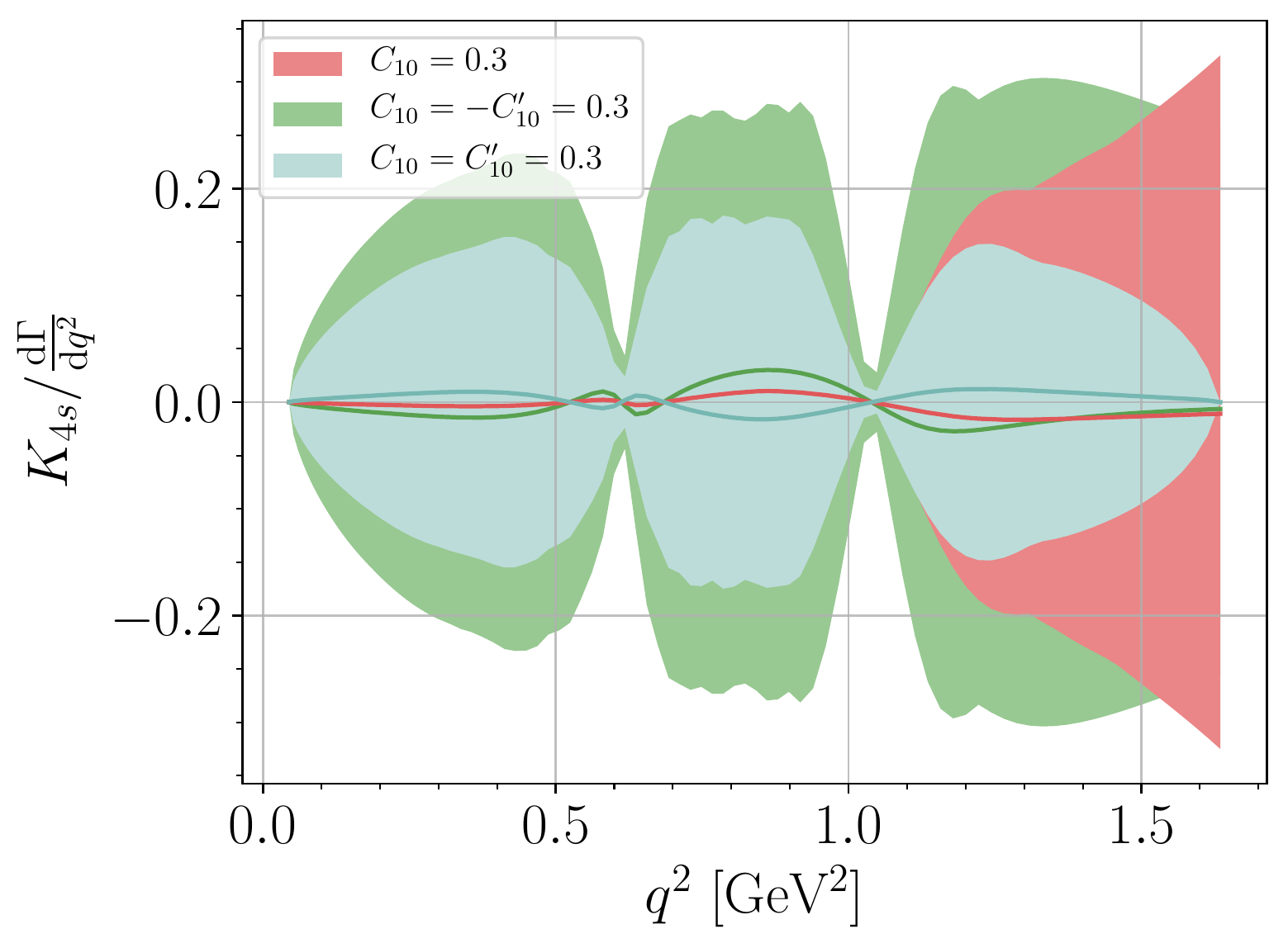}
  \includegraphics[width=0.48\textwidth]{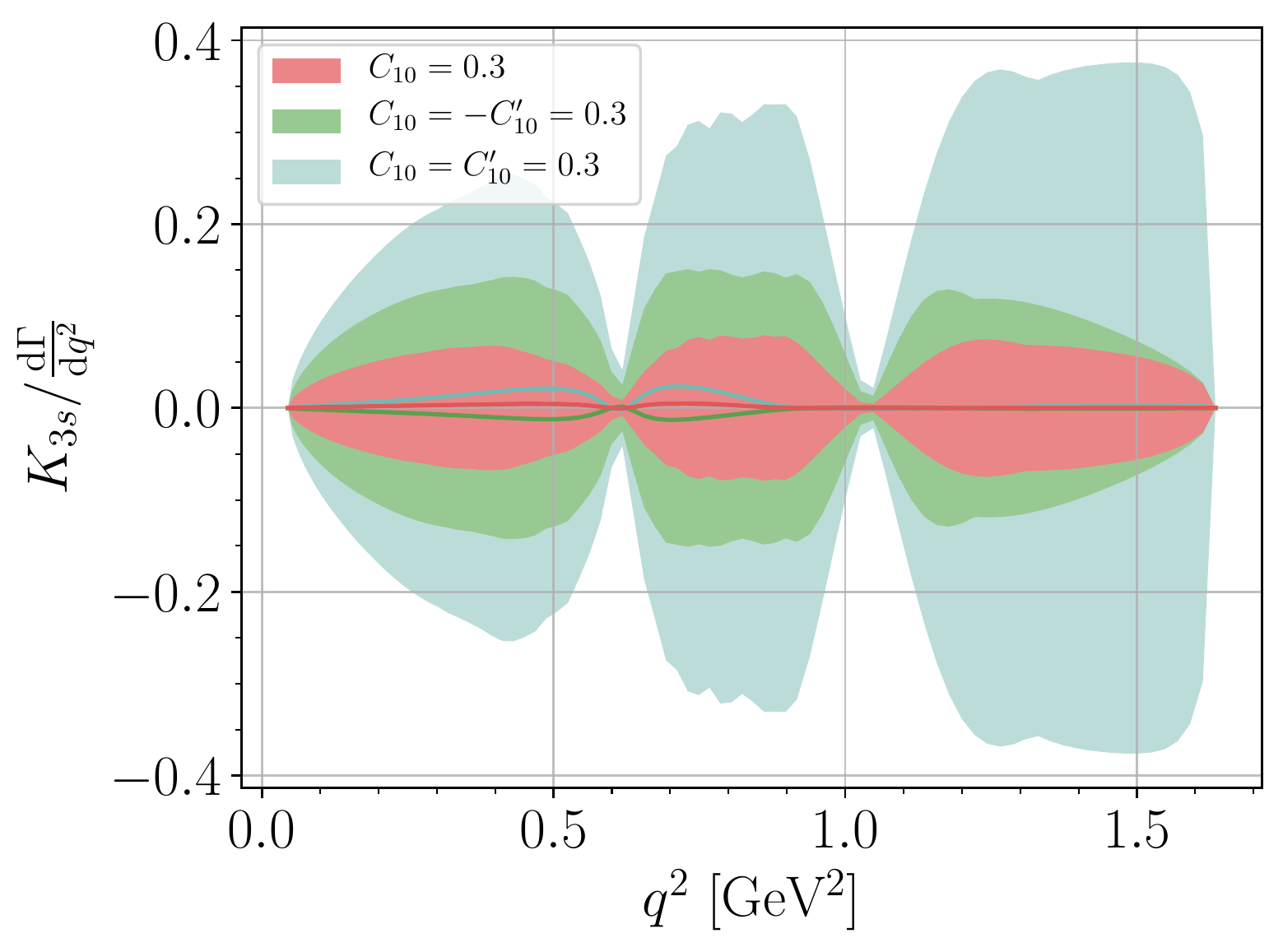}
\caption{The angular observables $K_{4s}$ (left) and $K_{3s}$ (right) normalized to the differential decay rate for $\Xi^+_c\to\Sigma^+(\to p\pi^0)\mu^+\mu^-$ decays in different NP scenarios for $C_{10}$ and $C^{\prime}_{10}$. The width of the bands stem predominantly from unknown strong phases. Both observables are clean SM null tests (\ref{eq:null}).}
\label{fig:i42}
\end{figure}

At zero recoil,  $K_{3s}= K_{1c}=0$, and
$K_{4s}  (\text{d} \Gamma/\text{d}q^2)^{-1} =-K_{2c} (\text{d} \Gamma/\text{d}q^2)^{-1}$ and in general finite, with the value dependent on the model  \cite{Hiller:2021zth}.
In Fig.~\ref{fig:i42} we show $K_{3s}$ (right panel) and $K_{4s}$ (left panel), normalized to the differential decay rate, for the same benchmarks 
with NP in $C^{(\prime)}_{10}$ as for 
 $A^{\ell\text{H}}_{\text{FB}}$ in Fig.~\ref{fig:afb_lH}.
While the different Wilson coefficient and form factor combinations of $K_{3s}$ and $K_{4s}$   only offer little qualitative complementarity compared to $A^{\ell\text{H}}_{\text{FB}}$, they do 
increase the statistics and enhance the sensitivity  in a global analysis.

\subsection{Baryonic dineutrino modes \label{sec:di}}

 Dineutrino modes induced by $c \to u \nu \bar \nu$ transitions are severely GIM suppressed and negligible in the SM  \cite{Burdman:2001tf}. 
Any observation hence signals NP, making them prime candidates for searches~\cite{Bause:2020xzj}. 
 The effective Hamiltonian reads
\begin{align}
    \mathcal{H}_{\text{eff}}=-\frac{4G_F}{\sqrt{2}}\sum_{ij}\left(C_L^{ij}Q_L^{ij}+C_R^{ij}Q_R^{ij} \right)\,,
\end{align}
with $C_{L,R}^{ij}$ negligible in the SM and 
\begin{align}
    Q_L^{ij}=(\overline{u}_L\gamma_\mu c_L)(\bar{\nu}_{Lj}\gamma^\mu \nu_{Li})\,,\quad Q_R^{ij}=(\overline{u}_R\gamma_\mu c_R)(\bar{\nu}_{Lj}\gamma^\mu \nu_{Li})\,.
\end{align}
Assuming  the absence of  light right-handed neutrinos,  only these two operators exist  for each combination of neutrino flavors $i,j$.

The self-analyzing four-body decays offer further opportunities for rare charm decays into dineutrinos.
Specifically, this concerns the decays  $\Xi_c^+\to\Sigma^+\,(\to p\pi^0) \nu \bar \nu   \,  ,  ~ \Xi_c^0\to\Lambda^0\,(\to p \pi^-)  \nu \bar \nu  \,,$ and 
$\Omega_c^0\to\Xi^0\,(\to \Lambda^{0} \pi^0)  \nu \bar \nu  $. 
The angular distribution is then given by
\begin{equation}
\frac{\text{d}^2\Gamma}{\text{d}q^2 \text{d} \! \cos\theta_\pi}=\int_{-1}^{1}\int_{0}^{2\pi}\frac{\text{d}^4\Gamma}{\text{d}q^2\text{d} \! \cos\theta_\pi \text{d} \! \cos\theta_\ell \text{d} \phi}\,\text{d} \phi\,\,\text{d} \! \cos\theta_\ell=2\,K_{1ss}+K_{1cc} + (2\,K_{2ss}+K_{2cc} )  \cos\theta_\pi \, ,
\label{eq:thetapi}
\end{equation} 
and 
 accessible without reconstructing the neutrinos \cite{Hiller:2021zth}. Note, here,  in $K_{1ss},\,K_{1cc},\,K_{2ss}$ and $K_{2cc}$ one has to replace $C_9=-C_{10}=\frac{4\pi}{\alpha_e}\,C^{ij}_L/2$ and $C^\prime_9=-C^{\prime }_{10}=\frac{4\pi}{\alpha_e}\,C^{ij}_R/2$, 
 skip the $m_\ell^2$ terms and incoherently sum the neutrino flavors $ij$. This gives rise to two independent observables, the differential decay rate and the hadronic forward-backward asymmetry
$A^{\text{H}}_{\text{FB}}$ (\ref{eq:afb_H}).

The hadronic forward-backward asymmetry $A_{\text{FB}}^{\text H}$  is obtained by integrating over the leptons phase space, and therefore can be measured
in  dineutrino modes. It reads
\begin{align}\label{eq:afbh_neutrino1}
A_{\text{FB}}^{\text H}(B_0 \to B_1 (\to B_2 \pi) \nu \bar \nu)= -\alpha\cdot \frac{(|C_L|^2- |C_R|^2)\,A(q^2)\sqrt{s_+  s_-}}{\vert C_L-C_R\vert^2\,B(q^2)s_+ +\vert C_L+C_R\vert^2\,C(q^2)s_-}
\end{align}
and  probes $C_R/C_L$. To ease notation here we omit the flavor indices.
In the limit $C_R=0$, the asymmetry becomes free of Wilson coefficients,
\begin{align}\label{eq:afbh_neutrino2}
A_{\text{FB}}^{\text H}(B_0 \to B_1 (\to B_2 \pi) \nu \bar \nu)= - \alpha \cdot \frac{A(q^2) \sqrt{s_+  s_-}}{B(q^2)s_+  +C(q^2)s_-} \, . 
\end{align}
The functions $A(q^{2}),B(q^{2})$ and $C(q^{2})$ 
contain form factors and kinematics  and can be seen in App.~\ref{app:hel_amp}.

Upper limits on the branching ratios can be  derived from a global   EFT-analysis~\cite{Bause:2020xzj}
\begin{equation} \label{eq:Brs-di}
\begin{split}
\mathcal{B}(\Xi_c^+\to \Sigma^+(\to p\pi^0)  \nu \bar \nu   ) \lesssim 3.9 \times10^{-5}\,,\\
\mathcal{B}(\Xi_c^0\to \Lambda^0(\to p\pi^-) \nu \bar \nu    ) \lesssim 3.6 \times10^{-6}\,,\\
\mathcal{B}(\Omega_c^0\to \Xi^0(\to \Lambda^0\pi^0)  \nu \bar \nu     )  \lesssim  7.1 \times10^{-5}\,,
\end{split}
\end{equation}
in agreement with results in Ref.~\cite{Golz:2021imq}.
The upper limits are stronger when assumptions on the lepton flavor are made~\cite{Bause:2020xzj}.
These are given in the next equation, with the first entry corresponding to 
charged lepton flavor conservation, and the even stronger one in parentheses assuming  lepton  universality:
\begin{equation} \label{eq:Brs-di-flavor}
\begin{split}
\mathcal{B}(\Xi_c^+\to \Sigma^+(\to p\pi^0)  \nu \bar \nu   ) \lesssim 1.1 \times10^{-5}\,, ~~(1.9 \times10^{-6})\,,\\
\mathcal{B}(\Xi_c^0\to \Lambda^0(\to p\pi^-) \nu \bar \nu    ) \lesssim 1.0 \times10^{-6}\,,~~(1.7 \times10^{-7})\,,\\
\mathcal{B}(\Omega_c^0\to \Xi^0(\to \Lambda^0\pi^0)  \nu \bar \nu     )  \lesssim  1.9 \times10^{-5} \, , ~~(3.4 \times10^{-6}) \, .
\end{split}
\end{equation}

\subsection{Polarized Charm baryons \label{sec:pol}}

Let us point out that studying decays of polarized charmed baryons introduces further null test observables on top of those in (\ref{eq:null}). We identify in total eight additional angular null tests probing leptonic axial vector currents, {\it i.e.,} $C_{10}^{(\prime)}$~\cite{deBoer:2018buv}, which are proportional to the initial $B_0$-polarization $P_{B_0}$ , 
\begin{align} \label{eq:null2}
K_{13},K_{16}, K_{18}, K_{20}, K_{22}, K_{24}, K_{26}, K_{28}|_{\text{SM}}  \simeq 0 \, ,
\end{align}
with the differential distribution \cite{Blake:2017une}  given in App.~\ref{app:angular_distr}.
 Among these,  $K_{13}, K_{22}$ and $K_{24}$ do not vanish  for $\alpha=0$ and hence can
be studied in the simpler three-body decays, including $\Lambda_c \to p 
\mu^+ \mu^-$. We note that  $K_{13} =-P_{B_0} K_{2c}$, anticipating that a study with polarized baryons
can access some of the null tests, here  $A_{\text{FB}}^{\ell\text{H}}$, that otherwise require  a self-analyzing four-body decay.  If in the future high luminosity sources of polarized $\Lambda_c$'s or other charmed baryons can be used, see \cite{deBoer:2017que} for LHC and $e^+ e^-$-collider possibilities,  we would like to come 
back to explore these observables and their NP reach in charm in more detail.

\section{Conclusions}\label{sec:conclusion}

We perform a full angular analysis of  baryonic $|\Delta c|=|\Delta u|=1$  four-body  decays with self-analyzing secondary baryon to  explore the BSM reach. 
We identify several  such modes: $\Xi_c^+\to\Sigma^+\,(\to p\pi^0)\ell^+\ell^-$, $\Xi_c^0\to\Lambda^0\,(\to p \pi^-)\ell^+\ell^-$ and  
$\Omega_c^0\to\Xi^0\,(\to \Lambda^{0} \pi^0)\ell^+\ell^-$,  with sizable decay parameter, see Table \ref{tab:list_channels}.
The full differential distribution of an unpolarized initial charm baryon (\ref{eq:angl_distr})
 features ten angular observables, seven more, and with different NP sensitivities, than the ones available with  three-body decays such as 
$\Lambda_c \to p \ell^+ \ell^-$.
 We point out   three new, clean null tests of the SM,  $K_{2c}, K_{3s}$ and $K_{4s}$ (\ref{eq:null}).
  Just like the leptonic  forward-backward asymmetry $A_{\text{FB}}^{\ell} \propto K_{1c}$ (\ref{eq:afb_l}), their  SM contribution can be safely neglected because the GIM-mechanism switches off axial-vector couplings of the  leptons,
  a feature previously exploited also for the  $D \to \pi \pi \ell^+ \ell^-$ angular distribution \cite{deBoer:2018buv}.
  We also find that the hadronic  forward-backward asymmetry $ A_{\text{FB}}^{\text{H}}$ (\ref{eq:afb_H}), although not a null test, can cleanly signal BSM physics in right-handed
  currents, $C_{7,9,10}^\prime$, illustrated in Fig.~\ref{fig:afb_H}.

The angular observables in four-body decays enable highly  diagnostic tests of BSM couplings.
Concrete analysis of the  NP sensitivity, see Section \ref{sec:nulltests},
shows that  already four observables,  the longitudinal polarization fraction $F_L$ (\ref{eq:fl}), together with the forward-backward asymmetries
$ A_{\text{FB}}^{\text{H}}$, $ A_{\text{FB}}^\ell$  and $A_{\text{FB}}^{\ell\text{H}}$ (\ref{eq:afb_lH}) allow  to pin down BSM Wilson coefficients in one go:

If there is sizable NP  only  in $F_L$, it is $C_7$.

If there is sizable NP only  in $F_L$ and $ A_{\text{FB}}^{\text{H}}$, it is $C_7^\prime$.

If there is sizable NP  only in $ A_{\text{FB}}^{\text{H}}$, it is $C_9^\prime$.

If there is sizable NP only  in $A_{\text{FB}}^{\ell\text{H}}$ and $ A_{\text{FB}}^{\text{H}}$, it is $C_{10}^\prime$.

If there is  sizable NP only  in $A_{\text{FB}}^{\ell}$ and $A_{\text{FB}}^{\ell\text{H}}$, it is $C_{10}$.

The price to pay for the self-analyzing  (quasi) four-body decay is the limitation to specific charm baryon modes; the suppression from the secondary baryon decays is modest since
branching ratios are at least 50 $\%$.  Experimental analysis is suitable for (advanced stages of)  high luminosity flavor factories LHCb~\cite{Cerri:2018ypt}, Belle II~\cite{Kou:2018nap}, BES III~\cite{Ablikim:2019hff}, and possible future machines~\cite{Charm-TauFactory:2013cnj,Abada:2019lih}.

Decays of polarized charmed baryon offer further GIM-based null tests (\ref{eq:null2}),
% from absence of $C_{10}^{(\prime)}$ in the SM,
some of which  persist in the simpler three-body decays. Their exploration should be pursued further if  in the future high luminosity sources of polarized $\Lambda_c$'s or other charmed baryons become available.
 
We note in passing that the hadronic forward-backward asymmetry is obtained by integrating over the leptons phase space, and therefore can be measured
in  dineutrino modes $\Xi_c^+\to\Sigma^+\,(\to p\pi^0) \nu \bar \nu   \,  ,  \Xi_c^0\to\Lambda^0\,(\to p \pi^-)  \nu \bar \nu $ and 
$\Omega_c^0\to\Xi^0\,(\to \Lambda^{0} \pi^0)  \nu \bar \nu $, briefly discussed in  Sec.~\ref{sec:di}, too.

We conclude that  rare decays of charm baryons contribute extensively to our endeavor to search for NP.
Dedicated computations of form factors  for $\Xi_c^+\to\Sigma^+$, $\Xi_c^0\to\Lambda^0$ and $\Omega_c^0\to\Xi^0$ transitions are desirable.
As anticipated  in~\cite{Golz:2021imq}, a  simultaneous fit of $|\Delta c|=|\Delta u|=1$   Wilson coefficients and resonance parameters is called for. The sensitivity of such a fit is enriched by the new presented observables in four-body baryon decays.

\section*{Acknowledgments}
We are happy to thank Dominik Mitzel for useful discussions. This work is supported by the \textit{Studienstiftung des Deutschen Volkes} (MG) and  in part by the \textit{Bundesministerium f\"ur Bildung und Forschung} (BMBF) under project number  05H21PECL2 (GH).

\appendix

\section{Helicity Amplitudes}\label{app:hel_amp}

Following \cite{Gutsche:2013pp} we introduce the contributions to the angular observables (\ref{eq:2})
\begin{align}
  \label{eq:1}
  \begin{split}
  &S^{m m'} = N^{2}\cdot\text{Re}\bigg[\mathcal{H}^{m}_{\frac{1}{2},t}\mathcal{H}^{\dagger m^{\prime}}_{\frac{1}{2},t} + \mathcal{H}^{m}_{-\frac{1}{2},t}\mathcal{H}^{\dagger m^{\prime}}_{-\frac{1}{2},t}\bigg],\\
  &S^{m m^{\prime}}_{P} = N^{2}\cdot\text{Re}\bigg[\mathcal{H}^{m}_{\frac{1}{2},t}\mathcal{H}^{\dagger m^{\prime}}_{\frac{1}{2},t} - \mathcal{H}^{m}_{-\frac{1}{2},t}\mathcal{H}^{\dagger m^{\prime}}_{-\frac{1}{2},t}\bigg],\\
  &U^{m m^{\prime}} = N^{2}\cdot\text{Re}\bigg[\mathcal{H}^{m}_{\frac{1}{2},1}\mathcal{H}^{\dagger m^{\prime}}_{\frac{1}{2},1} + \mathcal{H}^{m}_{-\frac{1}{2},-1}\mathcal{H}^{\dagger m^{\prime}}_{-\frac{1}{2},-1}\bigg],\\
  &P^{m m^{\prime}} = N^{2}\cdot\text{Re}\bigg[\mathcal{H}^{m}_{\frac{1}{2},1}\mathcal{H}^{\dagger m^{\prime}}_{\frac{1}{2},1} - \mathcal{H}^{m}_{-\frac{1}{2},-1}\mathcal{H}^{\dagger m^{\prime}}_{-\frac{1}{2},-1}\bigg],\\
  &L^{m m^{\prime}} = N^{2}\cdot\text{Re}\bigg[\mathcal{H}^{m}_{\frac{1}{2},0}\mathcal{H}^{\dagger m^{\prime}}_{\frac{1}{2},0} + \mathcal{H}^{m}_{-\frac{1}{2},0}\mathcal{H}^{\dagger m^{\prime}}_{-\frac{1}{2},0}\bigg],\\
  &L^{m m^{\prime}}_{P} = N^{2}\cdot\text{Re}\bigg[\mathcal{H}^{m}_{\frac{1}{2},0}\mathcal{H}^{\dagger m^{\prime}}_{\frac{1}{2},0} - \mathcal{H}^{m}_{-\frac{1}{2},0}\mathcal{H}^{\dagger m^{\prime}}_{-\frac{1}{2},0}\bigg],\\
  &I^{m m^{\prime}}_{1P} = \frac{N^{2}}{4}\text{Re}\bigg[
    \mathcal{H}^{m}_{\frac{1}{2},1}\mathcal{H}^{\dagger m^{\prime}}_{-\frac{1}{2},0} + \mathcal{H}^{m}_{-\frac{1}{2},0}\mathcal{H}^{\dagger m^{\prime}}_{\frac{1}{2},1} - \mathcal{H}^{m}_{\frac{1}{2},0}\mathcal{H}^{\dagger m^{\prime}}_{-\frac{1}{2},-1} - \mathcal{H}^{m}_{-\frac{1}{2},-1}\mathcal{H}^{\dagger m^{\prime}}_{\frac{1}{2},0}
    \bigg],\\
  &I^{m m^{\prime}}_{2} = \frac{N^{2}}{4}\text{Im}\bigg[
    \mathcal{H}^{m}_{\frac{1}{2},1}\mathcal{H}^{\dagger m^{\prime}}_{-\frac{1}{2},0}
    - \mathcal{H}^{m}_{-\frac{1}{2},0}\mathcal{H}^{\dagger m^{\prime}}_{\frac{1}{2},1}
    - \mathcal{H}^{m}_{\frac{1}{2},0}\mathcal{H}^{\dagger m^{\prime}}_{-\frac{1}{2},-1}+ \mathcal{H}^{m}_{-\frac{1}{2},-1}\mathcal{H}^{\dagger m^{\prime}}_{\frac{1}{2},0}
    \bigg],\\
  &I^{m m'}_{3} = \frac{N^{2}}{4}\text{Re}\bigg[
    \mathcal{H}^{m}_{\frac{1}{2},1}\mathcal{H}^{\dagger m^{\prime}}_{-\frac{1}{2},0}
    +\mathcal{H}^{m}_{-\frac{1}{2},0}\mathcal{H}^{\dagger m^{\prime}}_{\frac{1}{2},1}
    + \mathcal{H}^{m}_{\frac{1}{2},0}\mathcal{H}^{\dagger m^{\prime}}_{-\frac{1}{2},-1}
    + \mathcal{H}^{m}_{-\frac{1}{2},-1}\mathcal{H}^{\dagger m^{\prime}}_{\frac{1}{2},0}
    \bigg],\\
  &I^{m m'}_{4P} = \frac{N^{2}}{4}\text{Im}\bigg[
    \mathcal{H}^{m}_{\frac{1}{2},1}\mathcal{H}^{\dagger m^{\prime}}_{-\frac{1}{2},0}
    - \mathcal{H}^{m}_{-\frac{1}{2},0}\mathcal{H}^{\dagger m^{\prime}}_{\frac{1}{2},1}
    + \mathcal{H}^{m}_{\frac{1}{2},0}\mathcal{H}^{\dagger m^{\prime}}_{-\frac{1}{2},-1}
    - \mathcal{H}^{m}_{-\frac{1}{2},-1}\mathcal{H}^{\dagger m^{\prime}}_{\frac{1}{2},0}
    \bigg],
  \end{split}
\end{align}
in terms of helicity amplitudes $\mathcal{H}^{m}_{\lambda_{\Sigma},\lambda_\gamma}$, where $\lambda_{\Sigma}$ and $\lambda_\gamma$ denote the helicities of the $\Sigma$ baryon and the effective current $\gamma^* (\to \ell\ell)$, respectively. $\lambda_{\Sigma}$ can therefore assume values of $\pm\frac{1}{2}$, whereas $\lambda_\gamma$ takes values of $0,\,\pm1$, and we further distinguish $\lambda_\gamma = t$ in the $J_\gamma = 0$ case and  $\lambda_\gamma = 0$ in the $J_\gamma = 1$ case. The  superscript 
$m^{(\prime)}$ distinguishes between leptonic vector ($m^{(\prime)}=1$) and axial-vector ($m^{(\prime)}=2$) contributions. While the former receive contributions from
$C_7^{(\prime)}$ and $C_9^{(\prime)}$, the latter are induced by
$C_{10}^{(\prime)}$, hence vanish in the SM.
The helicity amplitudes $\mathcal{H}^{m}_{\lambda_{\Sigma}, \lambda_\gamma}$ are obtained by summing the contributions from individual hadronic matrix elements,
 $\mathcal{H}^{a,m}_{\lambda_{\Sigma}, \lambda_\gamma}$, hence $\mathcal{H}^{m}_{\lambda_{\Sigma}, \lambda_\gamma}=\sum_a \mathcal{H}^{a,m}_{\lambda_{\Sigma}, \lambda_\gamma}$. Here, $a=1,2$ correspond to dipole contributions  from 
$C_7^{(\prime)}$ and
$a=3,4$ to those from 4-fermion operators $C_{9}^{(\prime)}$ and $C_{10}^{(\prime)}$. Contributions  $a=1$ and $a=3$ are induced by quark-level vector currents, hence proportional to
$C+C^\prime$, whereas contributions  $a=2$ and $a=4$ are induced by quark-level axial-vector currents, hence $\propto C-C^\prime$.
Due to parity, flipping the helicities results in a minus sign for the amplitudes $a=2$ and $a=4$.
The amplitudes $\mathcal{H}^{1,2}_{\lambda_{\Sigma}, \lambda_\gamma}$ are then decomposed as
\begin{align}
\begin{split}
  &\mathcal{H}^{1}_{\lambda_{\Sigma},\lambda_\gamma} = \mathcal{H}^{{1},{1}}_{\lambda_{\Sigma},\lambda_\gamma} + \mathcal{H}^{{2},{1}}_{\lambda_{\Sigma},\lambda_\gamma} + \mathcal{H}^{{3},{1}}_{\lambda_{\Sigma},\lambda_\gamma} + \mathcal{H}^{{4},{1}}_{\lambda_{\Sigma},\lambda_\gamma},\\
  &\mathcal{H}^{1}_{-\lambda_{\Sigma},-\lambda_\gamma} = \mathcal{H}^{{1},{1}}_{\lambda_{\Sigma},\lambda_\gamma} - \mathcal{H}^{{2},{1}}_{\lambda_{\Sigma},\lambda_\gamma} + \mathcal{H}^{{3},{1}}_{\lambda_{\Sigma},\lambda_\gamma} - \mathcal{H}^{{4},{1}}_{\lambda_{\Sigma},\lambda_\gamma},\\
  &\mathcal{H}^{2}_{\lambda_{\Sigma},\lambda_\gamma} = \mathcal{H}^{{3},{2}}_{\lambda_{\Sigma},\lambda_\gamma} + \mathcal{H}^{{4},{2}}_{\lambda_{\Sigma},\lambda_\gamma},\\
  &\mathcal{H}^{2}_{-\lambda_{\Sigma},-\lambda_\gamma} = \mathcal{H}^{{3},{2}}_{\lambda_{\Sigma},\lambda_\gamma} - \mathcal{H}^{{4},{2}}_{\lambda_{\Sigma},\lambda_\gamma}\,.
\end{split}
\end{align}
 For convenience we give in the following a list of single contributions but stress that except for the different decay modes these equations are equivalent to Eqs.~(C3)-(C5) of Ref.~\cite{Golz:2021imq}. The helicity of the initial, charm baryon satisfies $\lambda_{\Xi_c} =-\lambda_{\Sigma}+\lambda_\gamma$.

\noindent
{$\lambda_{\Xi_c} = \frac{1}{2}$, $\lambda_{\gamma} = t$:}
\begin{align}
  \begin{split} 
  &\mathcal{H}^{1,1}_{-\frac{1}{2},t} = 0\,,\\
  &\mathcal{H}^{2,1}_{-\frac{1}{2},t} = 0\,,\\
  &\mathcal{H}^{3,1(2)}_{-\frac{1}{2},t} = (C_{9(10)}+C_{9\,(10)}^\prime)\frac{\sqrt{s_+}}{\sqrt{q^2}}f_0(q^2)(m_{\Xi_c}-m_\Sigma)\,,\\
  &\mathcal{H}^{4,1(2)}_{-\frac{1}{2},t} = (C_{9(10)}-C_{9\,(10)}^\prime)\frac{\sqrt{s_-}}{\sqrt{q^2}}g_0(q^2)(m_{\Xi_c}+m_\Sigma)\,,
  \end{split}
\end{align}
{$\lambda_{\Xi_c} = -\frac{1}{2}$, $\lambda_{\gamma} = 0$:}
\begin{align}
  \begin{split}
  &\mathcal{H}^{1,1}_{\frac{1}{2},0} = (C_7+C_7^\prime)\frac{2m_c}{\sqrt{q^2}}\sqrt{s_-}h_+(q^2)\,,\\
  &\mathcal{H}^{2,1}_{\frac{1}{2},0} = -(C_7-C_7^\prime)\frac{2m_c}{\sqrt{q^2}}\sqrt{s_+}\tilde{h}_+(q^2)\,,\\
  &\mathcal{H}^{3,1(2)}_{\frac{1}{2},0} = (C_{9(10)}+C_{9\,(10)}^\prime)\frac{1}{\sqrt{q^2}}\sqrt{s_-}f_+(q^2)(m_{\Xi_c}+m_\Sigma)\,,\\
  &\mathcal{H}^{4,1(2)}_{\frac{1}{2},0} = -(C_{9(10)}-C_{9\,(10)}^\prime)\frac{1}{\sqrt{q^2}}\sqrt{s_+}g_+(q^2)(m_{\Xi_c}-m_\Sigma)\,,
  \end{split}
\end{align}
{$\lambda_{\Xi_c} = \frac{1}{2}$, $\lambda_{\gamma} = 1$:}
\begin{align}
  \begin{split} 
  &\mathcal{H}^{1,1}_{\frac{1}{2},1} = \sqrt{2}(C_7+C_7^\prime)\frac{2m_c}{q^2}\sqrt{s_-}h_\bot(q^2)(m_{\Xi_c}+m_\Sigma)\,,\\
  &\mathcal{H}^{2,1}_{\frac{1}{2},1} = -\sqrt{2}(C_7-C_7^\prime)\frac{2m_c}{q^2}\sqrt{s_+}\tilde{h}_\bot(q^2)(m_{\Xi_c}-m_\Sigma)\,,\\
  &\mathcal{H}^{3,1(2)}_{\frac{1}{2},1} = \sqrt{2}(C_{9(10)}+C_{9\,(10)}^\prime)\sqrt{s_-}f_\bot(q^2)\,,\\
  &\mathcal{H}^{4,1(2)}_{\frac{1}{2},1} = -\sqrt{2}(C_{9(10)}-C_{9\,(10)}^\prime)\sqrt{s_+}g_\bot(q^2)\,.
  \end{split}
\end{align}

Using the helicity amplitudes we obtain for the contributions in  Eq.~\eqref{eq:1}
\begin{align}
  \begin{split}
  \label{eq:36}
  U^{11} =4N^{2}\cdot\bigg[\,& \bigg|(C_{7} + C_7^\prime)\,\frac{2m_{c}}{q^{2}}(m_{\Xi_c^+} + m_{\Sigma^+})\,h_{\perp} + (C_{9} + C_9^\prime)\,\,f_{\perp}\bigg|^{2}\cdot s_{-}\\
  + &\bigg|(C_{7} - C_7^\prime)\,\frac{2m_{c}}{q^{2}}(m_{\Xi_c^+} - m_{\Sigma^+})\,\tilde{h}_{\perp} + (C_{9} - C_9^\prime)\,\,g_{\perp}\bigg|^{2}\cdot s_{+} \, \bigg]\,,\\
  L^{11} =\frac{2N^{2}}{q^2}\cdot\bigg[\,&\bigg|(C_{7} + C_7^\prime)\,2m_{c}\,h_{+} + (C_{9} + C_9^\prime)\,(m_{\Xi_c^+} + m_{\Sigma^+})\,f_{+}\bigg|^{2}\cdot s_{-}\\
  + &\bigg|(C_{7} - C_7^\prime)\,2m_{c}\,\tilde{h}_{+} + (C_{9} - C_9^\prime)\,(m_{\Xi_c^+} -  m_{\Sigma^+})\,g_{+}\bigg|^{2}\cdot s_{+} \, \bigg]\,,\\
  U^{22} =4N^{2}\cdot\bigg[\,&\bigg|(C_{10} + C_{10}^\prime)\, f_{\perp}\bigg|^{2}\cdot s_{-} +\,\, \bigg|(C_{10} - C_{10}^\prime)\, g_{\perp}\bigg|^{2}\cdot s_{+}\, \bigg]\,,\\
  L^{22} =\frac{2N^{2}}{q^2}\cdot\bigg[\,&\bigg|(C_{10} + C_{10}^\prime)\, (m_{\Xi_c^+} + m_{\Sigma^+})\,f_{+}\bigg|^{2}\cdot s_{-} +\,\, \bigg|(C_{10} - C_{10}^\prime)\, (m_{\Xi_c^+} - m_{\Sigma^+})\,g_{+}\bigg|^{2}\cdot s_{+}\, \bigg]\,,\\
  S^{22} =\frac{2N^{2}}{q^2}\cdot\bigg[\, &\bigg|(C_{10} + C_{10}^\prime)\,(m_{\Xi_c^+} - m_{\Sigma^+})\,f_{0}\bigg|^{2}\cdot s_{+} +\,\, \bigg|(C_{10} - C_{10}^\prime)\,(m_{\Xi_c^+} + m_{\Sigma^+})\,g_{0}\bigg|^{2}\cdot s_{-}\, \bigg]\,,\\
  P^{12} = -8N^{2}\cdot\bigg[\,&\text{Re}\big((C_{7} - C_7^\prime)\,(C^{*}_{10} + C^{\prime *}_{10})\big)\,\frac{m_{c}}{q^{2}}(m_{\Xi_c^+} - m_{\Sigma^+})\,f_{\perp}\,\tilde{h}_{\perp}\\
  +\,&\text{Re}\big((C_{7} + C_7^\prime)\,(C^{*}_{10} - C^{\prime *}_{10})\big)\,\frac{m_{c}}{q^{2}}(m_{\Xi_c^+} + m_{\Sigma^+})\,g_{\perp}\,h_{\perp} \\+ &\,\text{Re}\big(C_{9} C^{*}_{10} - C_9^\prime C^{\prime *}_{10}\big)\, g_{\perp}\,f_{\perp}\bigg]\cdot\sqrt{s_{+}s_{-}}\,.
  \end{split}
\end{align}
Here, $N^2={\frac{G^{2}_{F}\alpha^{2}_{e}v\sqrt{\lambda(m^{2}_{\Xi^+_{c}},\,m^{2}_{\Sigma^+},\,q^{2})}}{3\cdot 2^{11}\pi^{5}m^{3}_{\Xi^+_{c}}}}$ with the Källén function $\lambda(a,\,b,\,c)=a^2+b^2+c^2-2\,(ab+ac+bc)$ and $s_{\pm} = (m_{\Xi^+_c} \pm m_{\Sigma^+})^2 - q^2$.  The contributions in Eq.~\eqref{eq:36} are those relevant to three-body decays
and have already been given in Ref.~\cite{Golz:2021imq}.
%\clearpage
The additional contributions that arise in the full,  four-body angular distribution~\eqref{eq:angl_distr} read
\begin{align}
  \begin{split}
  \label{eq:ang_with_wilsons}
  L_P^{11} =- \frac{4N^{2}}{q^2}\cdot&\text{Re}\bigg[ \bigg((C_7 + C^\prime_7)\,2 m_c h_+ +(C_9 + C^\prime_9)\, (m_{\Xi^+_c}+m_{\Sigma^+}) f_+\bigg) \\
  &\cdot  \bigg((C^*_7 - C^{*\prime}_7)\,2 m_c \tilde{h}_+ +(C^*_9 - C^{*\prime}_9)\, (m_{\Xi^+_c}-m_{\Sigma^+}) g_+\bigg)\bigg]\cdot\sqrt{s_+s_-}\,,\\
  P^{11} =- 8N^{2}\cdot&\text{Re}\bigg[ \bigg((C_7 +C^\prime_7)\,\frac{2 m_c}{q^2} h_\perp (m_{\Xi^+_c}+m_{\Sigma^+}) + (C_9 + C^\prime_9)\, f_\perp \bigg)\\
  &\cdot \bigg((C^*_7 -C^{*\prime}_7)\,\frac{2 m_c}{q^2} \tilde{h}_\perp (m_{\Xi^+_c}-m_{\Sigma^+}) + (C^*_9 - C^{*\prime}_9)\, g_\perp \bigg)
  \bigg]\cdot\sqrt{s_+s_-}\,,\\
  L_P^{22} = - \frac{4N^{2}}{q^2}\cdot&\bigg[(\vert C_{10}\vert^2 - \vert C^\prime_{10}\vert^2)\,f_+ g_+ (m_{\Xi^+_c}^2-m_{\Sigma^+}^2) \bigg]\cdot\sqrt{s_+s_-}, \\
  P^{22} = - 8N^{2}\cdot&\bigg[(\vert C_{10}\vert^2 - \vert C^\prime_{10}\vert^2)\,f_\perp g_\perp  \bigg]\cdot\sqrt{s_+s_-},\\
  U^{12} = \phantom{-} 4 N^2 \cdot & \bigg[ \bigg(\text{Re}((C_7+C_7^\prime)(C^*_{10}+C_{10}^{*\prime}))\,f_\perp h_\perp \frac{2 m_c}{q^2}(m_{\Xi_c^+}+m_{\Sigma^+}) \\
  &\,\,\,+ \text{Re}((C_9+C_9^\prime)(C^*_{10}+C_{10}^{*\prime}))\,f_\perp^2 \bigg)\cdot s_- \\
  &+\bigg(\text{Re}((C_7-C_7^\prime)(C^*_{10}-C_{10}^{*\prime}))\,g_\perp \tilde{h}_\perp \frac{2 m_c}{q^2}(m_{\Xi_c^+}-m_{\Sigma^+}) \\
  &\,\,\,+ \text{Re}((C_9-C_9^\prime)(C^*_{10}-C_{10}^{*\prime}))\,g_\perp^2 \bigg)\cdot s_+ \bigg],\\
  S^{22}_{P} = -\frac{4N^{2}}{q^{2}}\cdot & \bigg[\left(\vert C_{10}\vert^{2}-\vert C^{\prime}_{10}\vert^{2}\right)f_{0}g_{0}(m^{2}_{\Xi^{+}_{c}}-m^{2}_{\Sigma^{+}})\bigg]\cdot\sqrt{s_{+}s_{-}}.
  \end{split}
\end{align}

%\clearpage
The interference terms are given by
\begin{align}
\begin{split} \label{eq:inter}
I_{1P}^{11}= \phantom{2m_c} N^{2}\sqrt{\frac{2}{q^2}}\cdot &\bigg[ \text{Re}((C_7-C_7^\prime)(C^*_7+C_7^{*\prime}))\,\frac{4m_c^2}{q^2}\cdot\left(\tilde{h}_+h_\perp(m_{\Xi_c^+}+m_{\Sigma^+})-h_+\tilde{h}_\perp(m_{\Xi_c^+}-m_{\Sigma^+})\right) \\
  &+\text{Re}((C_9-C_9^\prime)(C^*_9+C_9^{*\prime}))\cdot\left(g_+f_\perp(m_{\Xi_c^+}-m_{\Sigma^+})-f_+g_\perp(m_{\Xi_c^+}+m_{\Sigma^+})\right) \\
  &+\text{Re}((C_9-C_9^\prime)(C^*_7+C_7^{*\prime}))\,2m_c\cdot\left(g_+h_\perp \frac{m_{\Xi_c^+}^2-m_{\Sigma^+}^2}{q^2}-h_+g_\perp\right)\\
  &+\text{Re}((C_7-C_7^\prime)(C^*_9+C_9^{*\prime}))\,2m_c\cdot\left(\tilde{h}_+f_\perp-f_+\tilde{h}_\perp \frac{m_{\Xi_c^+}^2-m_{\Sigma^+}^2}{q^2}\right)  \bigg]\cdot\sqrt{s_+s_-}\,,\\
I_{1P}^{22}=\quad N^2\,\sqrt{\frac{2}{q^2}}\cdot&\bigg[(\vert C_{10}\vert^2 - \vert C^\prime_{10}\vert^2)\cdot \left(f_\perp g_+ (m_{\Xi_c^+}-m_{\Sigma^+})-f_+g_\perp(m_{\Xi_c^+}+m_{\Sigma^+}) \right)\bigg]\cdot \sqrt{s_+s_-},\\
I_{2}^{11}=2m_c \, N^2\,\sqrt{\frac{2}{q^2}}\cdot&\bigg[\text{Im}((C_9+C_9^\prime)(C^*_7+C_7^{*\prime}))\cdot\left(f_\perp h_+ - f_+ h_\perp\,\frac{(m_{\Xi_c^+}+m_{\Sigma^+})^2}{q^2}\right)\cdot s_-\\
&-\text{Im}((C_9-C_9^\prime)(C^*_7-C_7^{*\prime}))\cdot\left(g_\perp \tilde{h}_+ - g_+ \tilde{h}_\perp\,\frac{(m_{\Xi_c^+}-m_{\Sigma^+})^2}{q^2}\right)\cdot s_+\bigg], \\
I_{2}^{22}= \quad0 , \quad\quad\quad\quad& \\
I_{3}^{12}= \phantom{2m_c} N^{2}\sqrt{\frac{2}{q^2}}\cdot &\bigg[ \text{Re}((C_7+C_7^\prime)(C^*_{10}+C_{10}^{*\prime}))\,m_c\cdot\left(h_+f_\perp+f_+ h_\perp \frac{(m_{\Xi_c^+}+m_{\Sigma^+})^2}{q^2}\right)\cdot s_- \\
  &-\text{Re}((C_7-C_7^\prime)(C^*_{10}-C_{10}^{*\prime}))\,m_c\cdot\left(\tilde{h}_+g_\perp+g_+ \tilde{h}_\perp \frac{(m_{\Xi_c^+}-m_{\Sigma^+})^2}{q^2}\right)\cdot s_+ \\
  &+\text{Re}((C_9+C_9^\prime)(C^*_{10}+C_{10}^{*\prime}))\cdot\left(f_+f_\perp(m_{\Xi_c^+}+m_{\Sigma^+})\right)\cdot s_-\\
  &-\text{Re}((C_9-C_9^\prime)(C^*_{10}-C_{10}^{*\prime}))\cdot\left(g_+g_\perp(m_{\Xi_c^+}-m_{\Sigma^+})\right)\cdot s_+\bigg]\,,\\ \\
I_{4P}^{12}= \phantom{2m_c} N^{2}\sqrt{\frac{2}{q^2}}\cdot &\bigg[\text{Im}((C_7+C_7^\prime)(C^*_{10}-C_{10}^{*\prime}))\,m_c\cdot\left(h_\perp g_+\frac{m_{\Xi_c^+}^2-m_{\Sigma^+}^2}{q^2}+h_+g_\perp\right)\\
&+\text{Im}((C_7-C_7^\prime)(C^*_{10}+C_{10}^{*\prime}))\,m_c\cdot\left(\tilde{h}_\perp f_+\frac{m_{\Xi_c^+}^2-m_{\Sigma^+}^2}{q^2}+\tilde{h}_+f_\perp\right)\\
&+\text{Im}((C_9+C_9^\prime)(C^*_{10}-C_{10}^{*\prime}))\frac{1}{2}\cdot\left(f_\perp g_+(m_{\Xi_c^+}-m_{\Sigma^+})+ f_+g_\perp (m_{\Xi_c^+}+m_{\Sigma^+}) \right)\\
&+\text{Im}((C_9-C_9^\prime)(C^*_{10}+C_{10}^{*\prime}))\frac{1}{2}\cdot\left(g_\perp f_+(m_{\Xi_c^+}+m_{\Sigma^+})+ g_+f_\perp (m_{\Xi_c^+}-m_{\Sigma^+}) \right)\bigg]\\&\cdot \sqrt{s_+s_-}\,.
\end{split}
\end{align}

All additional contributions   (\ref{eq:ang_with_wilsons}), (\ref{eq:inter})  except  $U^{12},I_{2}^{11}$ and $I_{3}^{12}$ are P-odd, that is, change sign for $C_i \leftrightarrow C_i^\prime$, and vanish for $C_i=C_i^\prime$.

In Eqs.~\eqref{eq:4},~\eqref{eq:afbh_neutrino1} and~\eqref{eq:afbh_neutrino2} the following $q^2$-dependent functions appear
\begin{align}
\begin{split}
A(q^2)&=2\,f_\perp g_\perp+f_+g_+\,\frac{m_{B_0}^2-m_{B_1}^2}{q^2}\,,\\
B(q^2)&=2\,g_\perp^2 + g^2_+\,\frac{\left(m_{B_0}-m_{B_1}\right)^2}{q^2}\,,\\
C(q^2)&=2\,f_\perp^2 + f^2_+\,\frac{\left(m_{B_0}+m_{B_1}\right)^2}{q^2}\,.\\
\end{split}
\end{align}

\section{Helicity amplitude description of $B_1\to B_2 \pi$}
\label{app:hadr_hel}

The secondary baryonic decay $B_1\to B_2 \pi$, here discussed for  $\Sigma^{+}\to p\pi^{0}$, can be parameterized by the sum, $\alpha_+$, and the difference, $\alpha_-$, of the  helicity amplitudes $h^{\Sigma}_{\lambda_{p}}(\lambda_{\Sigma})$ squared
\begin{equation}
  \label{eq:3}
  \alpha_{\pm} = \left|h^{\Sigma}_{\frac{1}{2}}\left(\lambda_{\Sigma}= \frac{1}{2}\right)\right|^{2} \pm \left|h^{\Sigma}_{-\frac{1}{2}}\left(\lambda_{\Sigma}=-\frac{1}{2}\right)\right|^{2}.
\end{equation}
The helicity amplitudes of non-leptonic baryon decays involving a spin-0 meson, here taken to be a pion, can be parametrized as
\begin{equation}
  \label{eq:nonleptbaryon}
  h^{\Sigma}_{\lambda_{p}}(\lambda_{\Sigma}) = G_{F}m^{2}_{\pi}\bar{u}_{p}(\lambda_{p})(A-B\gamma_{5})u_{\Sigma}(\lambda_{\Sigma}),
\end{equation}
where $A$ and $B$ are complex constants~\cite{Commins:1983ns} and $m_\pi$ the pion mass.
We compute the amplitude in the rest frame of the $\Sigma^{+}$ with the $z$-axis pointing in the direction of the proton momentum.
The spinors then take the form
\begin{align}
  \label{eq:nonleptbaryonspinors}
  \begin{split}
    u_{\Sigma}\left(p,\lambda_{\Sigma}=\pm\frac{1}{2}\right)&=\sqrt{2m_{\Sigma}}
    \begin{pmatrix}
      \chi_{\pm}\\0
    \end{pmatrix},\\
    \bar{u}_{p}\left(k,\lambda_{p}=\pm\frac{1}{2}\right)&=\sqrt{E_{p}+m_{p}}
    \left(\chi^{\dagger}_{\pm},\frac{\mp|\vec{k}|}{E_{p}+m_{p}}\chi^{\dagger}_{\pm}\right),
  \end{split}
\end{align}
where $p$  ($k$) denotes the four-momentum of the $\Sigma^{+}$ (proton),  with $p^0=m_\Sigma$, $\vert\vec{p}\vert=0$ and $E_p=\sqrt{\vert \vec{k}\vert^2+m_p^2}$ is  the energy of the proton, hence $k=(E_p, 0, 0, \vert \vec{k}\vert)^{\text{T}}$,
and $\chi_{+}= (1,0)^{\text{T}}$, $\chi_{-} = (0,1)^{\text{T}}$.
Plugging  the spinors into  (\ref{eq:nonleptbaryon}) and simplifying, we arrive at
\begin{align}
  \label{eq:hhelamps}
  \begin{split}
    h^{\Sigma}_{\frac{1}{2}}\left(\lambda_{\Sigma}=\frac{1}{2}\right)&=\sqrt{2m_{\Sigma}}G_{F}m^{2}_{\pi}(\sqrt{r_{+}}A+\sqrt{r_{-}}B),\\
    h^{\Sigma}_{-\frac{1}{2}}\left(\lambda_{\Sigma}=-\frac{1}{2}\right)&=\sqrt{2m_{\Sigma}}G_{F}m^{2}_{\pi}(\sqrt{r_{+}}A-\sqrt{r_{-}}B),
  \end{split}
\end{align}
with $r_{\pm}=\sqrt{E_{p}\pm m_{p}}$.
Using these helicity amplitudes we can express $\alpha_{\pm}$ as
\begin{align}
  \label{eq:alphaexpr}
  \begin{split}
    &\alpha_{+}=4G^{2}_{F}m^{4}_{\pi}m_{\Sigma}(r_{+}|A|^{2}+r_{-}|B|^{2}) \, , \\
    &\alpha_{-}=8G^{2}_{F}m^{4}_{\pi}m_{\Sigma}\sqrt{r_{+}r_{-}}\,\text{Re}(AB^{*}) \, , 
  \end{split}
\end{align}
and obtain for their ratio
\begin{equation}
  \label{eq:alphapdg}
  \frac{\alpha_{-}}{\alpha_{+}}=\frac{2\sqrt{\frac{r_{-}}{r_{+}}}\text{Re}(AB^{*})}{|A|^{2}+\frac{r_{-}}{r_{+}}|B|^{2}}=\alpha\, ,
\end{equation}
which corresponds to  the decay parameter $\alpha$ in~\cite{Zyla:2020zbs}.
We can therefore factorize $\alpha_{+}$ from the angular distribution and  use $\alpha_{+}=\mathcal{B}(\Sigma^{+}\to p\pi)$ and $\frac{\alpha_{-}}{\alpha_{+}}=\alpha$ to arrive at the expressions given in Sec.~\ref{Sec:angular-dist}.

\section{Angular distribution for polarized initial baryons}
\label{app:angular_distr}

Taking into account initial state polarization, the differential decay distribution depends on five angles and $q^2$ and reads
\begin{equation}
\begin{split}
\frac{\text{d}^{6}\Gamma}{\text{d}q^2\,\text{d}\vec{\Omega}} = \frac{3}{32\pi^{2}} \Big(
& \left(K_{1ss}\sin^2\theta_\ell+K_{1cc}\cos^2\theta_\ell+K_{1c}\cos\theta_\ell\right)  +  \\[-5pt]
& \left(K_{2ss}\sin^2\theta_\ell+K_{2cc}\cos^2\theta_\ell+K_{2c}\cos\theta_\ell\right)\cos\theta_\pi +  \\
&\left(K_{3sc}\sin\theta_\ell\cos\theta_\ell+K_{3s}\sin\theta_\ell\right)\sin\theta_\pi\sin\left(\phi_c+\phi_\ell\right) +  \\
& \left(K_{4sc}\sin\theta_\ell\cos\theta_\ell+K_{4s}\sin\theta_\ell\right)\sin\theta_\pi\cos\left(\phi_c+\phi_\ell\right) +  \\
& \left(K_{11}\sin^2\theta_\ell+K_{12}\cos^2\theta_\ell+K_{13}\cos\theta_\ell\right)\cos\theta_c +  \\
& \left( K_{14}\sin^2\theta_\ell+K_{15}\cos^2\theta_\ell+K_{16}\cos\theta_\ell\right)\cos\theta_\pi \cos\theta_c +  \\
& \left(K_{17}\sin\theta_\ell\cos\theta_\ell+K_{18}\sin\theta_\ell\right)\sin\theta_\pi\cos\left(\phi_c+\phi_\ell\right)\cos\theta_c  +  \\
& \left(K_{19}\sin\theta_\ell\cos\theta_\ell+K_{20}\sin\theta_\ell\right)\sin\theta_\pi\sin\left(\phi_c+\phi_\ell\right) \cos\theta_c +  \\
& \left(K_{21}\cos\theta_\ell\sin\theta_\ell+K_{22}\sin\theta_\ell\right)\sin\phi_\ell \sin\theta_c +  \\
& \left(K_{23}\cos\theta_\ell\sin\theta_\ell+K_{24}\sin\theta_\ell\right)\cos\phi_\ell  \sin\theta_c +  \\
& \left(K_{25}\cos\theta_\ell\sin\theta_\ell+K_{26}\sin\theta_\ell\right)\sin\phi_\ell\cos\theta_\pi  \sin\theta_c +  \\
& \left(K_{27}\cos\theta_\ell\sin\theta_\ell+K_{28}\sin\theta_\ell\right)\cos\phi_\ell\cos\theta_\pi  \sin\theta_c  +  \\
& \left(K_{29}\cos^2\theta_\ell+K_{30}\sin^2\theta_\ell\right)\sin\theta_\pi\sin\phi_c  \sin\theta_c +  \\
& \left(K_{31}\cos^2\theta_\ell+K_{32}\sin^2\theta_\ell\right)\sin\theta_\pi\cos\phi_c  \sin\theta_c +  \\
& \left(K_{33}\sin^2\theta_\ell \right) \sin\theta_\pi\cos\left(2\phi_\ell+\phi_c\right) \sin\theta_c  +  \\
& \left(K_{34}\sin^2\theta_\ell \right) \sin\theta_\pi\sin\left(2\phi_\ell+\phi_c\right)  \sin\theta_c  \Big)~.
\end{split}
\label{eq:34terms}
\end{equation} 

$K_{11},\,K_{12},\,K_{13},\,K_{21},\,K_{22},\,K_{23}$ and $K_{24}$ survive in the limit $\alpha=0$ of which $K_{13},\,K_{22}$ and $K_{24}$ are null tests. The first four lines are identical to Eq.~\eqref{eq:angl_distr} with $\phi_c+\phi_\ell=\phi$. $\phi_c$ and $\theta_c$ are new angles related to the initial state polarization, see Ref.~\cite{Blake:2017une} for details.

\end{document}